\documentclass{emulateapj}
\usepackage{apjfonts}
\usepackage{psfrag}
\newcommand{\myemail}{psharma@pppl.gov}
\newcommand{\ba}{\begin{eqnarray}}
\newcommand{\ea}{\end{eqnarray}}
\newcommand{\be}{\begin{equation}}
\newcommand{\ee}{\end{equation}}
\newcommand{\Par}{\parallel}
\newcommand{\Perp}{\perp}
\newcommand{\grad}{\nabla}

\slugcomment{{\sc Submitted to ApJ}: August 23, 2005}

\shorttitle{Simulations of the MRI in a Collisionless Plasma}
\shortauthors{Sharma et al.}

\begin{document}

\title{Shearing Box Simulations of the MRI in a Collisionless Plasma}

\author{Prateek Sharma, Gregory W. Hammett}
\affil{Princeton Plasma Physics Laboratory, Princeton University,
    Princeton, NJ 08543}
\email{\myemail, hammett@pppl.gov}

\author{Eliot Quataert}
\affil{Astronomy Department, University of California,
    Berkeley, CA 94720}
\email{eliot@astron.berkeley.edu}

\and

\author{James M. Stone}
\affil{Princeton University Observatory, Princeton University,
   Princeton, NJ 08540}
\email{jstone@astro.princeton.edu}

\begin{abstract}

We describe local shearing box simulations of turbulence driven by the
magnetorotational
instability (MRI) in a collisionless plasma.  Collisionless effects may
be important in radiatively inefficient accretion flows, such as near the
black hole in the Galactic Center. The MHD version of ZEUS
is modified to evolve an anisotropic pressure tensor.  A fluid closure
approximation is used to calculate heat conduction along magnetic field lines.
The anisotropic pressure tensor provides a qualitatively new mechanism
for transporting angular momentum in accretion flows (in addition to
the Maxwell and Reynolds stresses).  We estimate limits on the pressure
anisotropy due to pitch angle scattering by kinetic instabilities.  
Such instabilities provide an effective ``collision'' rate in a 
collisionless plasma and lead to more MHD-like dynamics. We find that the 
MRI leads to efficient growth of the magnetic field in a collisionless 
plasma, with saturation amplitudes comparable to those in MHD.  In the
saturated state, the anisotropic stress is comparable to the Maxwell
stress, implying that the rate of angular momentum transport may be
moderately enhanced in a collisionless plasma.

\end{abstract}


\section{Introduction}

Following the seminal work of \citet{bal91}, numerical simulations
have demonstrated that magnetohydrodynamic (MHD) turbulence initiated
by the magnetorotational instability (MRI) is an efficient mechanism
for transporting angular momentum in accretion disks (e.g., Hawley,
Gammie, \& Balbus 1995, hereafter \citet{haw95}; see Balbus \& Hawley 1998, 
for a review). For a broad class of astrophysical accretion
flows, however, the MHD
assumption is not directly applicable.  In particular, in radiatively
inefficient accretion flow (RIAF) models for accretion onto compact
objects, the accretion proceeds via a hot, low density, collisionless
plasma with the proton temperature larger than the electron
temperature \citep[see][for reviews]{nar98, qua03}.  In order to
maintain such a two-temperature flow the plasma must be collisionless,
and there are many cases where the Coulomb mean-free path is many orders
of magnitude larger than the system size.
Motivated by the application to RIAFs, this paper studies the
nonlinear evolution of the MRI in a collisionless plasma, focusing on
local simulations in the shearing box limit.

Quataert, Dorland, \& Hammett (2001; hereafter \citet{qua02}) and Sharma,
Hammett, \& Quataert (2003; hereafter \citet{sha03}) showed that the linear
dynamics of the MRI in a collisionless plasma can be quite different
from that in MHD.  The maximum growth rate is a factor of $\sim 1.7$
larger and, perhaps more importantly, the fastest growing modes can
shift to much longer wavelengths, giving direct amplification of long
wavelength modes.  Dynamical instability exists even when the magnetic
tension forces are
negligible because of the anisotropic pressure response in a
collisionless plasma.  In related work using Braginskii's anisotropic
viscosity,
\citet{bal04} has termed this modification of the MRI in the low
collisionality limit the ``magnetoviscous'' instability
\citep[see also][]{isl05}.

In this paper, we are interested in simulating the dynamics of a
collisionless plasma on length-scales ($\sim$ disk height) and
time-scales ($\sim$ orbital period) that are very large compared to
the microscopic plasma scales (such as the Larmor radius and the
cyclotron period). Since the ratio of the size of the accretion flow
to the proton Larmor radius is $\sim 10^{8}$ for typical RIAF models,
direct particle methods such as PIC (particle in a cell), which need to
resolve both of these scales, are computationally challenging and
require simulating a reduced range of scales.
Instead we use a fluid-based method to describe the
large-scale dynamics of a collisionless plasma (``kinetic MHD,''
described in \S 2). The key differences with respect to MHD are that the
pressure is a tensor rather than a scalar, anisotropic with respect to
the direction of the local magnetic field, and that there are heat
fluxes along magnetic field lines (related to Landau damping and
wave-particle interactions).  
The
drawback of our fluid-based method is, of course, that there is no
exact expression for the heat fluxes if only a few fluid moments are
retained in a weakly collisional plasma (the
``closure problem'').  We use results from Snyder, Hammett, \& Dorland
(1997; hereafter \citet{sny97}) who
have derived approximations for the heat fluxes in terms of nonlocal parallel
temperature and magnetic field gradients. These heat flux expressions
can be shown to be equivalent to multi-pole Pad\'e approximations to the
$Z$-function involved in Landau damping.  This approach can be
shown to converge as more fluid moments of the distribution function are kept
{\citep{Hammett93}},
just as an Eulerian kinetic algorithm converges as more grid points in
velocity space are kept.
These fluid-based methods
have been applied with reasonable success to modeling collisionless
turbulence in fusion plasmas,
generally coming within a factor of 2 of more complicated kinetic
calculations in strong turbulence regimes
\citep[e.g.,][]{dimits00,par94,Hammett93,Scott05},
though there can be larger differences in weak turbulence regimes
\citep[see][and references therein]{Hammett93,dimits00}.
The simulations we report on here use an even simpler local
approximation to the heat flux closures than those derived in \citet{sny97}.
While not exact, these closure approximations allow one to begin to
investigate kinetic effects with relatively fast modifications
of fluid codes.

In a collisionless plasma the magnetic moment $\mu$
is an adiabatic invariant.
Averaged over velocity space, this leads to conservation of
$\langle \mu \rangle =p_\Perp/(\rho B)$.
As a result, pressure anisotropy with
$p_\Perp>p_\Par$ is created as the MRI amplifies the magnetic field in
the accretion flow.  This pressure anisotropy creates an anisotropic
stress (like a viscosity!) which can be as important for angular momentum
transport as the magnetic stress (as we show below).  The pressure
anisotropy cannot, however, grow without bound because high frequency
waves and kinetic microinstabilities feed on the free energy in the
pressure anisotropy, effectively providing an enhanced rate of
collisions that limit the pressure tensor anisotropy (leading to
more MHD-like dynamics in a collisionless plasma).  We capture
this physics by using a subgrid model to restrict the allowed 
amplitude of the pressure anisotropy.  This subgrid model (described
in \S 2.3) is based on existing linear and nonlinear studies of
instabilities driven by pressure anisotropy \citep[e.g.,][]{has69,gar97}. 

The remainder of this paper is organized as follows. In the next
section ($\S 2$) we describe Kulsrud's formulation of kinetic MHD
(KMHD) and our closure model for the heat fluxes in a collisionless
plasma. We also include a linear analysis of the MRI in the presence
of a background pressure anisotropy and describe limits on the 
pressure anisotropy set by kinetic instabilities.  In $\S 3$ we
describe our modifications to the ZEUS code to model kinetic
effects. \S 4 presents our primary results on the nonlinear evolution
of the MRI in a collisionless plasma. In \S 5 we discuss these
results, their astrophysical implications, and future work required to
understand the global dynamics of collisionless accretion disks.

\section{Governing Equations}
In the limit that all fluctuations of interest are at scales larger
than the proton Larmor radius and have frequencies much smaller than
the proton cyclotron frequency, a collisionless plasma can be
described by the following magnetofluid equations
\citep[e.g.,][]{kul83,sny97}: 
\ba
\label{eq:MHD1}
\frac{\partial \rho}{\partial t} & + & \nabla \cdot \left(\rho {\bf
V}\right)=0,
\\
\label{eq:MHD2}
\rho \frac{\partial {\bf V}}{\partial t} & + & \rho\left({\bf V} \cdot
\nabla\right)
{\bf V}= \frac{\left(\nabla \times {\bf B}\right) \times {\bf B}}{4\pi} 
- \nabla \cdot {\bf P} + {\bf F_g},\\
\label{eq:MHD3}
\frac{\partial {\bf B}}{\partial t} &=& \nabla \times \left({\bf V} \times
{\bf
B}\right), \\
\label{eq:MHD4}
 {\bf P}&=& p_{\Perp} {\bf I} + \left(p_{\Par}- p_{\Perp}\right){\bf
\hat{b}\hat{b}} = p_\Perp {\bf I} + {\bf \Pi}, \ea where $\rho$ is the
mass density, ${\bf V}$ is the fluid velocity, ${\bf B}$ is the
magnetic field, ${\bf F_g}$ is the gravitational force, ${\bf
\hat{b}}={\bf B}/|{\bf B}|$ is a unit vector in the direction of the
magnetic field, and ${\bf I}$ is the unit tensor. In
equation~(\ref{eq:MHD3}) an ideal Ohm's law is used, neglecting
resistivity.  In equation (\ref{eq:MHD4}), ${\bf P}$ is the pressure
tensor with different perpendicular~($p_{\Perp}$) and
parallel~($p_{\Par}$) components with respect to the background
magnetic field, and ${\bf \Pi} = {\bf \hat{b} \hat{b}} (p_\Par -
p_\Perp)$ is the anisotropic stress tensor. 
(Note that ${\bf \Pi}$ is not traceless in the convention used here.)
${\bf P}$ should in
general be a sum over all species but in the limit where ion dynamics
dominate and electrons just provide a neutralizing background, the
pressure can be interpreted as the ion pressure. This is the case for
hot accretion flows in which $T_p \gg T_e$.

The exact pressures $p_\Par$ and $p_\Perp$ can be rigorously
determined by taking moments of the drift kinetic equation
\citep{kul83}, 
\ba
\label{eq:DKE}
\nonumber
\frac{\partial f_s}{\partial t} &+& (v_\Par {\bf \hat{b} + v_E}) \cdot
\nabla f_s \\
\nonumber
&+& \left [ -{\bf \hat{b}} \cdot \frac{D {\bf v_E}}{Dt} - \mu
{\bf \hat{b}} \cdot \nabla B + \frac{e_s}{m_s} \left ( E_\Par +
\frac{F_{g\Par}}{e_s} \right ) \right ] \frac{\partial f_s}{\partial
v_\Par} \\
&=& C(f_s) 
\ea 
which is the asymptotic expansion of the Vlasov
equation for the distribution function $f_s({\bf x}, \mu, v_\Par, t)$
for species `$s$' with mass $m_s$ and charge $e_s$
in the limit $\rho_s/L \ll 1$, $\omega/\Omega_s \ll
1$, where $\rho_s$ and $\Omega_s$ are the gyroradius and
gyrofrequency, respectively. In equation (\ref{eq:DKE}), ${\bf
v_E}=c({\bf E} \times {\bf B})/B^2$ is the perpendicular drift
velocity, $\mu=({\bf v_\Perp} -{\bf v_E})^2/2B$ is the magnetic moment
(a conserved quantity in the absence of collisions), $F_{g \Par}$ is the
component of the gravitational force parallel to the direction of the
magnetic field, and $D/Dt=\partial/\partial t + (v_\Par {\bf
\hat{b}+v_E}) \cdot \nabla$ is the particle Lagrangian derivative 
in configuration space.
The fluid velocity ${\bf V}= {\bf v_E + \hat{b}} u_\Par$, so
the ${\bf E} \times {\bf B}$ drift is determined by the perpendicular
component of equation~(\ref{eq:MHD2}).  Other drifts such as grad B,
curvature, and gravity $\times {\bf B}$ drifts are higher order in the
drift kinetic MHD ordering
and do not appear in this equation.
In equation (\ref{eq:DKE}), $C(f_s)$ is the collision operator to
allow for generalization to collisional regimes. Collisions can also
be used to mimic rapid pitch angle scattering due to high frequency
waves that break $\mu$ invariance.  The parallel electric field is
determined by $E_\| = \sum_s (e_s/m_s) {\bf \hat{b}} \cdot \grad \cdot 
{\bf P}_s/\sum_s (n_s e_s^2/m_s) $ \citep{kul83}{}, which insures
quasineutrality.

Separate equations of state for the parallel and perpendicular
pressures can also be obtained from the moments of the drift kinetic
equation \citep[][]{che56}.  Neglecting the collision term these are:
\ba
\label{eq:CGL1}
\rho B \frac{D}{Dt} \left( \frac{p_\Perp}{\rho B} \right) &=& - \nabla \cdot
{\bf  q_\Perp} - q_\Perp \nabla \cdot {\bf \hat{b}}, \\
\label{eq:CGL2}
\frac{\rho^3}{B^2} \frac{D}{Dt} \left( \frac{p_\Par B^2}{\rho^3}
\right) &=& -\nabla \cdot {\bf q_\Par} + 2 q_\Perp \nabla \cdot
{\bf\hat{b}}, \ea where $D/Dt = \partial/\partial t + {\bf V} \cdot
\nabla$ is the fluid Lagrangian derivative and ${\bf q_{\Par,\perp}} =
q_{\Par, \Perp} {\bf \hat{b}}$ are the heat fluxes (flux of $p_\Par$ and
$p_\Perp$) parallel to the magnetic field.  The equation for
the magnetic moment density $\rho \langle \mu \rangle 
=p_\Perp/B$ can be written in a conservative form: \be
\label{eq:mucon}
\frac{\partial}{\partial t} \left( \frac{p_\Perp}{B} \right) + \nabla
\cdot \left(\frac{p_\Perp}{B} {\bf V} \right) = - \nabla \cdot \left(
\frac{q_\Perp}{B}{\bf \hat{b}} \right), \ee  If the heat fluxes are
neglected (called the CGL or double adiabatic limit), as the magnetic
field strength ($B$) increases, $p_\Perp$ increases ($p_\Perp \propto
\rho B$), and $p_\Par$ decreases ($p_\Par \propto
\rho^3/B^2$). Integrating equation~(\ref{eq:mucon}) over a finite
periodic (even a shearing periodic) box shows that $\langle p_\Perp/B
\rangle$ is conserved, where $\langle \rangle$ denotes a volume
average. This implies that even when $q_{\Par,\Perp} \ne 0$, $p_\Perp$
increases in a volume averaged sense as the magnetic energy in the box
increases. This means that that for a collisionless plasma, pressure
anisotropy $p_\Perp >(<)\,p_\Par$ is created as a natural consequence
of processes that amplify (reduce) $B$. This pressure anisotropy is
crucial for understanding magnetic field amplification in
collisionless dynamos.

To solve the set of equations (\ref{eq:MHD1}-\ref{eq:MHD4}),
(\ref{eq:CGL1}-\ref{eq:CGL2}) in a
simple fluid based formalism, we require expressions for $q_\Par$ and
$q_\Perp$ in terms of lower order moments. No simple exact expressions for
$q_\Par$ and $q_\Perp$ exist for nonlinear collisionless plasmas. Although
simple, the double adiabatic or CGL approximation (where $q_\Par=
q_\Perp=0$) does not capture key kinetic effects such as Landau
damping.  In the moderately collisional limit ($\rho_i <$ mean free
path $<$ system size), where the distribution function is not very
different from a local Maxwellian, one can use the Braginskii 
equations~\citep{bra65} to describe anisotropic transport 
\citep[see][for astrophysical applications]{bal00,bal04}.
However, in the hot RIAF regime, the
mean free path is often much larger than the system size and the Braginskii
equations are not formally applicable, though they are still useful 
as a qualitative indication of the importance of kinetic effects.
The collisional limit of the kinetic MHD equations can be shown to recover
the dominant anisotropic heat flux and viscosity tensor of
Braginskii~\citep{sny97}.  The local approximation to kinetic MHD that we
use here leads to equations that are similar in form to Braginskii MHD, but
with separate dynamical equations for parallel and perpendicular 
pressures.  We also add models for enhanced pitch angle scattering by
microinstabilities, which occur at very small scales and high frequencies
beyond the range of validity of standard kinetic MHD.
\footnote{This would also be needed when using Braginskii equations, 
because they are not necessarily well posed in situations where
the anisotropic stress tensor can drive arbitrarily small scale
instabilities\citep{sch05}.}

Hammett and collaborators have developed approximate fluid closures
(called Landau fluid closure) for collisionless plasmas
\citep[][]{ham90,ham92,sny97} that capture kinetic effects such
Landau damping. \citet{sny97} give the resulting expressions for
parallel heat fluxes ($q_\Par$, $q_\Perp$) to be used in equations
(\ref{eq:CGL1}) and (\ref{eq:CGL2}).  Landau closures are based on
Pad\'e approximations to the full kinetic plasma dispersion function that
reproduce the correct asymptotic behavior in both the $\omega/k_\Par
c_\Par \gg 1$ and $\omega/k_\Par c_\Par \ll 1$ regimes (and provide a good
approximation in between), where $\omega$ is
the angular frequency, $k_\Par$ is the wavenumber parallel to the
magnetic field, and $c_\Par=\sqrt{p_\Par/\rho}$ is the parallel thermal
velocity of the particles.  In Fourier space, the linearized heat fluxes can be
written as ~\citep[equations (39) \& (40) in][]{sny97} 
\ba
q_\Par= && -\sqrt{\frac{8}{\pi}} \rho_0  c_{\Par 0} 
\frac{ik_\Par \left( p_\Par/\rho \right)}{|k_\Par|}, \\
\nonumber
q_\Perp= && -\sqrt{\frac{2}{\pi}}
\rho_0  c_{\Par 0} 
   \frac{ik_\Par \left (p_\Perp/\rho \right)}{|k_\Par|} \\
&& +
\sqrt{\frac{2}{\pi}}  c_{\Par 0} \frac{p_{\Perp 0}}{B_0} 
   \left ( 1-  \frac{p_{\Perp 0}}{p_{\Par 0} }
   \right) \frac{ik_\Par B}{|k_\Par|}.  
\ea
where $0$ subscripts indicate equilibrium quantities.  Real space
expressions are somewhat more cumbersome and are given by
convolution integrals
\ba
  \label{eq:nonlocal1}
q_\Par = &-& \left( \frac{2}{\pi}\right)^{3/2} n_0  c_{\Par 0 } 
    \int_0^\infty \delta z^\prime 
    \frac{T_\Par(z+z^\prime)-T_\Par(z-z^\prime)}{z^\prime}, \\
\nonumber
q_\Perp = &-& \left( \frac{2}{\pi^3} \right)^{1/2} n_0 c_{\Par 0} 
    \int_0^\infty \delta z^\prime
\frac{T_\Perp(z+z^\prime)-T_\Perp(z-z^\prime)}{z^\prime} \\
\nonumber
   &+& \left( \frac{2}{\pi^3} \right)^{1/2} 
c_{\Par 0} \left(1- \frac{p_{\Perp 0}}{p_{\Par 0}}  \right) 
   \frac{  p_{\Perp 0} }{B_0} \times \\
\label{eq:nonlocal2}
&& \int_0^\infty
\delta z^\prime \frac{B(z+z^\prime)-B(z-z^\prime)}{z^\prime},
\ea
where
$n_0$ is the number density, $T_\Par=p_\Par/n$ and
$T_\Perp=p_\Perp/n$ are the parallel and perpendicular temperatures,
and $z^\prime$ is the spatial variable along the magnetic field
line. \citet{sha03} have shown that these fluid closures for the heat fluxes
accurately reproduce the kinetic linear Landau damping rate for all
MHD modes (slow, Alf\'ven, fast and entropy modes). The growth rate of the
MRI using the Landau closure model is also very similar to that
obtained from full kinetic theory. As noted in the introduction, in
addition to reproducing linear modes/instabilities, Landau fluid
closures have also been used to model turbulence in fusion plasmas
with reasonable success.

These closure approximations were originally developed for turbulence
problems in fusion energy devices with a strong guide magnetic field,
where the parallel dynamics is essentially linear and FFTs could be
easily used to quickly evaluate the Fourier expressions above.  In
astrophysical problems with larger amplitude fluctuations and tangled
magnetic fields, evaluation of the heat fluxes become somewhat more
complicated.  One could evaluate the convolution expressions, equations
(\ref{eq:nonlocal1}) and (\ref{eq:nonlocal2}) (with some modest
complexity involved in writing a subroutine to integrate along magnetic
field lines), leading to a code with a computational time $T_{cpu}
\propto N_x^3 N_\Par$, where $N_x^3$ is the number of spatial grid points
and $N_\Par$ is the number of points kept in the integrals along field
lines.  (In some cases, it may be feasible to map the fluid quantities
to and from a field-line following coordinate system so that FFTs can
reduce this to $T_{cpu} \propto N_x^3 \log N_\Par$. )  While this is more
expensive than simple MHD where $T_{cpu} \propto N_x^3$, it could still
represent a savings over a direct solution of the drift kinetic
equation, which would require $T_{cpu} \propto N_x^3 N_{v_\Par}
N_{v_\Perp}$, where $N_{v_\Par} N_{v_\Perp}$ is the number of grid points
for velocity space.  \footnote{On the other hand, an effective hyperdiffusion
operator in velocity space may reduce the velocity resolution
requirements, and recent direct kinetic simulations of turbulence in
fusion devices have found that often one does not need very high
velocity resolution.  This may make a direct solution of the drift
kinetic equation tractable for some astrophysical kinetic MHD problems.
Furthermore, a direct solution of the drift kinetic equation involves
only local operations, and thus is somewhat easier to parallelize than
the convolution integrals.}

As a first step for studying kinetic effects, in this paper
we pick out a characteristic wavenumber $k_L$ that represents the scale
of collisionless damping and use a local 
approximation for the heat fluxes in Fourier space, with a
straightforward assumption about the nonlinear generalization: \ba
\label{eq:qpar1}
q_\Par &=& -\sqrt{\frac{8}{\pi}} \rho  c_{\Par} 
\frac{\nabla_\Par \left( p_\Par/\rho \right)}{k_L}, \\
\label{eq:qperp1}
\nonumber
q_\Perp &=& -\sqrt{\frac{2}{\pi}} \rho  c_{\Par} 
\frac{\nabla_\Par \left( p_\Perp/\rho \right)}{k_L} \\
&& + \sqrt{\frac{2}{\pi}} c_\Par  p_\Perp 
\left( 1- \frac{p_\Perp}{p_\Par} \right) \frac{\nabla_\Par B}{k_LB}.  \ea
Note that this formulation of the heat flux is analogous to a
Braginskii heat conduction along magnetic field lines.  For
linear modes with $|k_\Par| \sim k_L$, these approximations will of
course agree with kinetic theory as well as the Pad\'e approximations
shown in \citep{sny97}.  One can think of $k_L$ as approximately
controlling the heat conduction rate, though this does not necessarily
affect the resulting Landau damping rate of a mode in a monotonic way,
since this sometimes exhibits impedance matching behavior.  I.e.,
some modes are weakly damped in both the small and large (isothermal)
heat conduction limits.  We will vary $k_L$ to investigate the
sensitivity of our results to this parameter.

\subsection{Linear Modes}
\label{sec:linmodes}

Since pressure anisotropy arises as a consequence of magnetic field
amplification in a collisionless plasma, it is of interest to repeat
the linear analysis of the collisionless MRI done previously in
\citet{qua02}, but 
with a background pressure anisotropy ($p_{\Par 0} \ne p_{\Perp 0}$). 
We consider the simple
case of a vertical magnetic field.  This analysis provides a useful guide to
understanding some of our numerical results.

We linearize equations (\ref{eq:MHD1})-(\ref{eq:MHD4}) for a
differentially rotating disk (${\bf V_0}= R \Omega(R) {\bf
\hat{\phi}}$) with an anisotropic pressure about a uniform subthermal
vertical magnetic field (${\bf B_0} = B_z{\bf \hat{z}}$).  We assume
that the background (unperturbed) plasma is described by a
bi-Maxwellian distribution ($p_{\Par 0} \neq p_{\Perp 0}$). We also
assume that the perturbations are axisymmetric, of the form exp[$-i
\omega t +i {\bf k \cdot x}$] with ${\bf k} = k_R {\bf \hat{R}} + k_z
{\bf \hat{z}}$. Writing $\rho=\rho_0+\delta \rho$, ${\bf B} = {\bf
B_0} + {\bf \delta B}$, $p_\Perp = p_{\Perp 0} + \delta p_\Perp$,
$p_\Par = p_{\Par 0} + \delta p_\Par$, working in cylindrical
coordinates and making a $|k|R \gg 1$ assumption, the 
linearized versions of equations
(\ref{eq:MHD1})-(\ref{eq:MHD3}) become: \ba
\label{eq:lin1}
\omega \delta \rho &=& \rho_0 {\bf k} \cdot \delta{\bf v}, \\
\label{eq:lin2}
\nonumber
-i \omega \rho_0 \delta v_R &-& \rho_0 2\Omega \delta v_{\phi}= -\frac{i
k_R}{4 \pi}B_z \delta B_z \\ 
&+& i k_z \left( \frac{B_z}{4 \pi} - \frac{(p_{\Par 0} -p_{\Perp 0})}{B_z} \right) \delta B_R - i k_R \delta p_{\Perp}, \\
\label{eq:lin3}
-i \omega \rho_0 \delta v_{\phi} &+& \rho_0 \delta v_R \frac{\kappa^2} {2
\Omega} = i k_z \left( \frac{B_z}{4 \pi}-\frac{(p_{\Par 0}-p_{\Perp 0})}{B_z}
\right) \delta B_\phi, \\
\label{eq:lin4}
-i \omega \rho_0 \delta v_z &=& -i k_R \left( p_{\Par 0}-p_{\Perp 0} \right)
\frac{\delta B_R}{B_z} - i k_z \delta p_{\Par}, \\
\label{eq:lin5}
\omega \delta B_R &=& - k_z B_z \delta v_R, \\
\label{eq:lin6}
\omega \delta B_{\phi} &=& - k_z B_z \delta v_{\phi} - \frac{i k_z
B_z}{\omega} \frac{d \Omega}{d \ln R} \delta v_R, \\
\label{eq:lin7}
\omega \delta B_z &=& k_R B_z \delta v_R, \ea where $\kappa^2=4\Omega^2
+ d\Omega^2/d \ln R$ is the epicyclic frequency.
Equations~(\ref{eq:lin1})-(\ref{eq:lin7}) describe the linear modes of
a collisionless disk with an initial pressure anisotropy about a
vertical magnetic field.  This corresponds to the $\theta=\pi/2$ case
of \citet{qua02}, but with an anisotropic initial pressure. Equations
(\ref{eq:lin2}) \& (\ref{eq:lin3}) show that an initial
anisotropic pressure modifies the Alfv\'en wave characteristics, so we
expect a background pressure anisotropy to have an important effect on
the MRI.  One way of interpreting equations (\ref{eq:lin2}) \&
(\ref{eq:lin3}) is that $p_\Perp>p_\Par$ ($p_\Par>p_\Perp$) makes the
magnetic fields more (less) stiff; as a result, this will shift the
fastest growing MRI mode to larger (smaller) scales.

The linearized equations for the parallel and perpendicular pressure
response are 
given by equations (33) and (34) in \citet{sha03}. We present them here for the
sake of completeness.  \ba
\label{eq:ppar}
-i \omega \delta p_\Par &+& p_{\Par 0} i {\bf k} \cdot {\bf \delta v} + i k_z
q_\Par + 2 p_{\Par 0} i k_z \delta v_z  = 0, \\
\label{eq:pperp}
-i \omega \delta p_\Perp &+& 2 p_{\Perp 0} i {\bf k} \cdot {\bf \delta
v} + i k_z q_\Perp - p_{\Perp 0} i k_z \delta v_z = 0, 
\ea 
where the
heat fluxes can be expressed in terms of lower moments using \ba
\label{eq:qperp}
\nonumber
q_\Perp &=& -\sqrt{\frac{2}{\pi}} c_{\Par 0}  \frac{i k_z}{|k_z|}
(\delta p_\Perp -  c_{\Par 0}^2 \delta \rho)
+ \sqrt{\frac{2}{\pi}}  c_{\Par 0} p_{\Perp 0} \times  \\
&& \left( 1
- \frac{p_{\Perp 0}}{p_{\Par 0}}  \right) \frac{i k_z}{|k_z|}
\frac{\delta B}{B_z}  , \\
\label{eq:qpar}
q_\Par &=& -\sqrt{\frac{8}{\pi}} c_{\Par 0}  
    \frac{i k_z}{|k_z|} (\delta p_\Par
-  c_{\Par 0} ^2 \delta \rho),
\ea
where $c_{\Par 0}=\sqrt{p_{\Par 0}/\rho_0}$ 
and $\delta B=|{\bf \delta B}|$.

\begin{figure}
\epsscale{.80}
\plotone{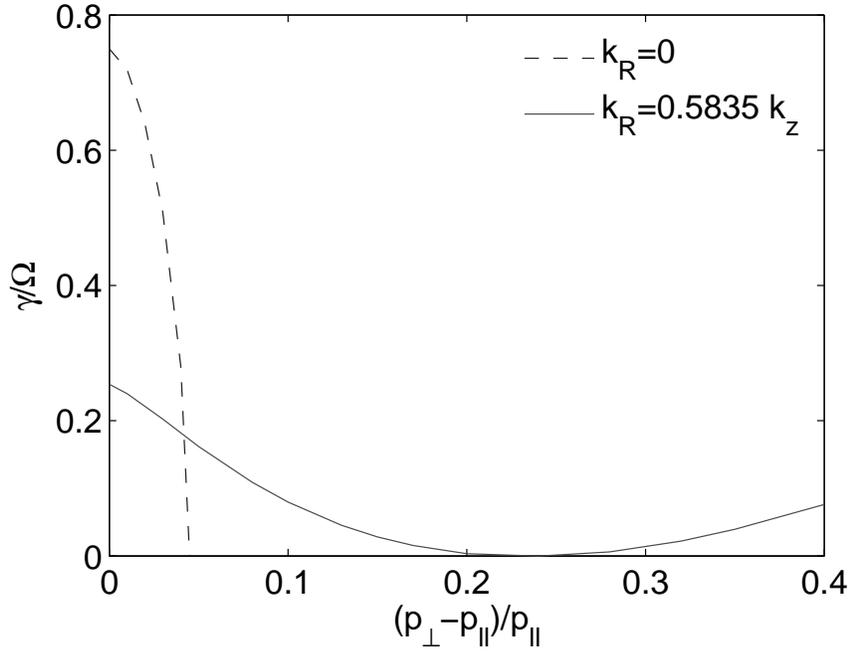}
\caption{Normalized growth rate ($\gamma/\Omega$) of the MRI versus
normalized pressure anisotropy, $(p_\Perp-p_\Perp)/p_\Par$ for
$\beta=100$, $k_z V_{Az}/\Omega=\sqrt{15/16}$, and two different
$k_R$'s.  Note that even a small anisotropy can stabilize the fastest
growing MRI mode.  The growth at large pressure anisotropy for $k_R
\ne 0$ is due to the mirror mode.
\label{fig:figure1}}
\end{figure}

Figure~\ref{fig:figure1} shows the MRI growth rate as a function of
pressure anisotropy for two values of $k_R$ for $\beta=100$. This
figure shows that the fastest growing MHD mode ($k_R=0$) is stabilized
for $(p_{\Perp 0}-p_{\Par 0})/p_{\Par 0} \sim 4/\beta$; modes with
$k_R \ne 0$ modes require larger anisotropy for stabilization.  For
$\beta \gg 1$, these results highlight that only a very small pressure
anisotropy is required to stabilize the fastest growing MRI modes.
Growth at large pressure anisotropies in Figure \ref{fig:figure1} for
$k_R \neq 0$ mode is because of the mirror instability that is
discussed below.  The physical interpretation of the stabilization of
the MRI in Figure \ref{fig:figure1} is that as the pressure anisotropy
increases ($p_{\Perp 0}>p_{\Par 0}$), the field lines effectively
become stiffer and modes of a given $k$ can be stabilized (though
longer wavelength modes will still be unstable).
In a numerical simulation in which the pressure anisotropy is allowed
(unphysically, as we see in \S 2.2) to grow without bound as the magnetic 
field grows, this
effect is capable of stabilizing all of the MRI modes in the
computational domain at very small amplitudes (see
Fig. \ref{fig:figure6} discussed in \S4).

\subsection{Isotropization of the Pressure Tensor in Collisionless
Plasmas}

Pressure anisotropy ($p_\perp \ne p_\Par$) is a source of free energy
that can drive instabilities which act to isotropize the pressure,
effectively providing an enhanced ``collision'' rate in a
collisionless plasma \citep[e.g.,][]{gar97}.  In order to do so, the
instabilities must break magnetic moment conservation, and thus must
have frequencies comparable to the cyclotron frequency and/or parallel
wavelengths comparable to the Larmor radius.  Because of the large
disparity in timescales between $\mu$-breaking microinstabilities and
the MRI ($\omega_{micro}/\Omega \sim 10^8$), one can envision the
microinstabilities as providing a ``hard wall'' limit on the pressure
anisotropy: once the pressure anisotropy exceeds the threshold value
where microinstabilities are driven and cause rapid pitch angle
scattering, the pressure anisotropy nearly instantaneously reduces the
anisotropy to its threshold value (from the point of view of the
global disk dynamics).  
In this section we review the relevant
instabilities that limit the pressure anisotropy in high $\beta$
collisionless plasmas -- these are the firehose, mirror, and ion
cyclotron instabilities.  We then discuss how we have implemented these
estimated upper bounds on the pressure anisotropy in our numerical
simulations.

\subsubsection{Maximum anisotropy for $p_\Par > p_\Perp$}

Plasmas with $p_\Par > p_\Perp$ can be unstable to the firehose
instability, whose dispersion relation for parallel propagation
is given by equation (2.12) of \citet{ken67}:
\be 
\omega^2 -
\omega \Omega_i k_\Par^2 \rho_i^2 + \Omega_i^2 k_\Par^2 \rho_i^2
\left( 1 - \frac{p_\Perp}{p_\Par} - \frac{2}{\beta_\Par} \right)=0,
\ee where $\beta_\Par=8\pi p_\Par/B^2$, $\rho_i$ is the ion Larmor
radius, $\Omega_i$ is the ion cyclotron frequency, and $k_\Par$ is the
wavenumber parallel to the local magnetic field direction. Solving for
$\omega$ gives \be \omega = k_\Par^2\rho_i^2 \frac{\Omega_i}{2} \pm i
k_\Par c_{\Par 0} 
\left(1 - \frac{p_\Perp}{p_\Par} - \frac{2}{\beta_\Par} -
\frac{k_\Par^2 \rho_i^2}{4} \right)^{1/2} \ee For long wavelengths,
the firehose instability requires $p_\Par>p_\Perp+B^2/4\pi$ and is
essentially an Alfv\'en wave destabilized by the pressure anisotropy.
The maximum growth rate occurs when $k_\Par^2 \rho_i^2 = 2
(1-p_\Perp/p_\Par-2/\beta_\Par)$ and is given by
$\Omega_i(1-p_\Perp/p_\Par-2/\beta_\Par)$. We use an upper limit on
$p_\Par>p_\Perp$ corresponding to $1-p_\Perp/p_\Par-2/\beta_\Par<1/2$,
which is an approximate condition for the growth of modes that will
violate $\mu$ conservation and produce rapid pitch angle scattering.

\subsubsection{Maximum anisotropy for $p_\Perp > p_\Par$}
\label{sec:mirror_scattering}

For $p_\Perp > p_\Par$ there are two instabilities that act to
isotropize the pressure, the mirror instability and the ion cyclotron
instability \citep[e.g.,][]{gar97}.  A plasma is unstable to the
mirror instability when $p_\Perp/p_\Par - 1 > 1/\beta_\Perp$,
although as discussed below only for somewhat larger anisotropies is
magnetic moment conservation violated.  Formally, a plasma with any
nonzero pressure anisotropy can be unstable to the ion cyclotron
instability \citep[][]{sti92}.  However, there is an effective threshold
given by the requirement that the unstable modes grow on a timescale
comparable to the disk rotation period.

Equations (43$^\prime$) \& (44$^\prime$) of \citet{has69} give the
wavenumber for the fastest growing mirror mode \ba
&&\frac{k_\Par}{k_\Perp}=\sqrt{\frac{(D-1)}{4}}, \\ &&k_\Perp \rho_i =
\sqrt{\frac{(D-1)}{6}}, \ea where $D=\beta_\Perp (p_\Perp/p_\Par-1)$,
$\beta_\Perp=8\pi p_\Perp/ B^2$. To estimate the pressure anisotropy
at which $\mu$ conservation is broken and thus pitch angle scattering
is efficient, we calculate $D$ for which $k_\Par \rho_i \sim k_\Perp
\rho_i \sim 1$. This implies $D \approx 7$ or that $\mu$ conservation
fails (and pitch angle scattering occurs) if the pressure anisotropy
satisfies \be
\label{eq:mirror_thresh}
\frac{p_\Perp}{p_\Par} - 1 > \frac{7}{\beta_\Perp}. \ee

The ion cyclotron instability can be also be excited when
$p_\Perp>p_\Par$. Gary and collaborators have analyzed the ion
cyclotron instability in detail through linear analysis and numerical
simulations. \citet{gar97} and \citet{gar94} calculate the pressure
anisotropy required for a given growth rate $\gamma$ relative to the
ion cyclotron frequency $\Omega_i$ \be
\label{eq:gary_thresh}
\frac{p_\Perp}{p_\Par} - 1 > \frac{S^\prime}{\beta_\Par^p} \ee where
$S^\prime=0.35$ and $p=0.42$ are fitting parameters quoted in equation
(2) of \citet{gar94} for $\gamma/\Omega_i=10^{-4}$.  Moreover, for
$\gamma \ll \Omega_i$ the threshold anisotropy depends only very
weakly on the growth rate $\gamma$.  
As a result, equation
(\ref{eq:gary_thresh}) provides a reasonable estimate of the pressure
anisotropy required for pitch angle scattering by the ion cyclotron
instability to be important on a timescale comparable to the disk
rotation period.

\subsection{Pressure Anisotropy Limits}
\label{sec:anisotropy}

Motivated by the above considerations, we require that the pressure
anisotropy satisfy the following inequalities in our simulations (at
each grid point and for all time steps): \ba 
\label{eq:pitch1}
&&
\frac{p_\Perp}{p_\Par}-1 +\frac{2}{\beta_\Par} >\frac{1}{2},\\
\label{eq:pitch2}
&&\frac{p_\Perp}{p_\Par}-1<\frac{2\xi}{\beta_\Perp},\\
\label{eq:pitch3}
&&\frac{p_\Perp}{p_\Par}-1<S \left(\frac{2}{\beta_\Par}\right)^{1/2},
\ea where $S$ and $\xi$ are constants described below.  It is
important to note that the fluid-based kinetic theory utilized in this
paper can correctly reproduce the existence and growth rates of the
firehose and mirror instabilities (though not the ion cyclotron
instability).\footnote{The double adiabatic limit ($q_\Perp = q_\Par =
0$) predicts an incorrect threshold and incorrect growth rates for the
mirror instability \citep[e.g.,][]{sny97}. Thus it is important to
use the heat flux models described in \S 2 to capture the physics of
the mirror instability.}  However, it can only do so for long
wavelength perturbations that conserve $\mu$. The relevant modes for
pitch angle scattering occur at the Larmor radius scale, which is very small
in typical accretion flows and is
unresolved in our simulations.  For this reason we must impose limits
on the pressure anisotropy and cannot simultaneously simulate the MRI
and the relevant instabilities that limit the pressure anisotropy.
The algorithm to impose the pressure anisotropy limits is explained in
Appendix~\ref{app:pthresh}. 

In equation (\ref{eq:pitch2}), the parameter $\xi$ determines
the threshold anisotropy above which the mirror instability leads to
pitch angle scattering. A value of $\xi = 3.5$ was estimated in
\S~\ref{sec:mirror_scattering}.  We take this as our fiducial value
but for comparison also describe calculations with $\xi = 0.5$,
which corresponds to the marginal state for the mirror instability.
We compare both models because the saturation of the mirror
instability is not well understood, particularly under the conditions
appropriate to a turbulent accretion disk.  Equation (\ref{eq:pitch3})
is based on the pitch angle scattering model used by~\citet{bir01} for
simulations of magnetic reconnection in collisionless plasmas;
following them we choose $S=0.3$. Equation (\ref{eq:pitch3})
with $S=0.3$ gives results which are nearly identical 
(for the typical range of $\beta$ studied here) to the pressure
anisotropy threshold for the ion cyclotron instability discussed in \S
\ref{sec:mirror_scattering} (eq.[\ref{eq:gary_thresh}]).

In our simulations we find that for typical calculations if $\xi =
0.5$ then equation (\ref{eq:pitch2}) (the ``mirror instability'')
dominates the isotropization of the pressure tensor while if $\xi =
3.5$ then equation (\ref{eq:pitch3}) (the ``ion cyclotron
instability'') dominates.  We also find that our results are
insensitive to the form of the $p_\Par > p_\Perp$ threshold
(eq. [\ref{eq:pitch1}]); e.g., simulations with $1 - p_\Perp/p_\Par <
2/\beta_\Par$ (the marginal state of the firehose mode) instead of
equation (\ref{eq:pitch1}) give nearly identical results.  Future
fully kinetic simulations of the mirror, firehose, and ion cyclotron
instabilities will be useful for calibrating the pitch angle
scattering models used here.

\section{Numerical Methods}

In this section we discuss the shearing box equations that we solve
numerically and the modifications made to ZEUS to include kinetic
effects.

\subsection{Shearing Box}

The shearing box is based on a local expansion of the tidal forces in
a reference frame corotating with the disk (see HGB for details). A
fiducial radius $R_0$ in the disk is picked out and the analysis is
restricted to a local Cartesian patch such that $L_x,L_y,L_z \ll R_0$
(where $x=r-R_0$, $y=\phi$ and $z=z$). In this paper only the radial
component of gravity is considered and buoyancy effects are ignored.
We also assume a Keplerian rotation profile. With these
approximations, the equations of Landau MHD in the shearing box are: \ba
\label{eq:SB1}
\frac{\partial \rho}{\partial t} &+& \nabla \cdot (\rho {\bf V}) = 0, \\
\nonumber
\frac{\partial {\bf V}}{\partial t}  &+& {\bf V} \cdot \nabla {\bf V} =
-\frac{1}{\rho} \nabla \left( p_\Perp + \frac{B^2}{8\pi} \right)
+ \frac{{\bf B} \cdot \nabla {\bf B}}{4\pi \rho} \\
\label{eq:SB2}
&-& \frac{1}{\rho} \nabla \cdot {\bf \Pi}
- 2 {\bf \Omega} \times {\bf V} + 3 \Omega^2 x {\bf \hat{x}}, \\
\label{eq:SB3}
\frac{\partial {\bf B}}{\partial t} &=& \nabla \times ({\bf V} \times
{\bf B}), \\ 
\nonumber 
\frac{\partial p_\Par}{\partial t} &+&
\nabla \cdot (p_\Par {\bf V}) + \nabla \cdot {\bf q_\Par} + 2 p_\Par
{\bf \hat{b}} \cdot \nabla {\bf V} \cdot {\bf \hat{b}} - 2q_\Perp
\nabla \cdot {\bf \hat{b}} \\
\label{eq:SB4}
&=& -\frac{2}{3} \nu_{eff} (p_\Par - p_\Perp), \\
\nonumber
\frac{\partial p_\Perp}{\partial t}  &+& \nabla \cdot (p_\Perp {\bf V})
+ \nabla \cdot {\bf  q_\Perp}  + p_\Perp \nabla \cdot {\bf V}
- p_\Perp {\bf \hat{b}} \cdot
\nabla {\bf V} \cdot {\bf \hat{b}} \\
\label{eq:SB5}
&+& q_\Perp \nabla \cdot {\bf \hat{b}}
= -\frac{1}{3} \nu_{eff} (p_\Perp - p_\Par), \\
\label{eq:SB6}
q_\Par &=& - \rho \kappa_\Par \nabla_\Par \left( \frac{p_\Par}{\rho}\right), \\
\label{eq:SB7}
q_\Perp &=& - \rho \kappa_\Perp \nabla_\Par \left(
\frac{p_\Par}{\rho}\right) + \kappa_m {\bf B} \cdot \nabla B, \ea
where ${\bf q_\Par} = q_\Par {\bf \hat{b}}$ and ${\bf q_\Perp} =
q_\Perp {\bf \hat{b}}$ are the heat fluxes parallel to the magnetic
field, $\nu_{eff}$ is the effective pitch-angle scattering rate
(including microinstabilities, see \S2.3 and Appendix \ref{app:pthresh}),
$\kappa_\Par$ and
$\kappa_\Perp$ are the coefficients of heat conduction, and $\kappa_m$
is the coefficient in $q_\Perp$ due to parallel gradients in the
strength of magnetic field~\citep[][]{sny97}. The $\kappa_m$ component
of $q_\Perp$ that arises because of parallel magnetic field gradients
is important for correctly recovering the saturated state for the mirror
instability in the fluid limit, where (in steady state) $q_{\Par,
\Perp} \approx 0$ implies that $T_\Par$ is constant along the field
line, and $T_\Perp$ and magnetic pressure are anticorrelated.

Given our closure models, the coefficients for the heat flux are given
by \ba
\label{eq:kappa_par}
\kappa_\Par &=& \frac{8 p_\Par}{\rho} \frac{1}{\sqrt{8 \pi \frac{p_\Par}{\rho}}
k_L + (3\pi-8)\nu_{eff} }, \\
\label{eq:kappa_perp}
\kappa_\Perp &=& \frac{p_\Par}{\rho} \frac{1}{\sqrt{\frac{\pi}{2}\frac{p_\Par}{\rho}}
k_L + \nu_{eff} }, \\
\label{eq:kappa_mag}
\kappa_m &=& \left(1-\frac{p_\Perp}{p_\Par}\right)
\frac{p_\Perp}{B^2} \kappa_\Perp, \ea where $k_L$ is the parameter
that corresponds to a typical wavenumber characterizing Landau damping
(see \S2).  We consider several values of $k_L$ to study the effect of
Landau damping on different scales. In particular, we consider
$k_L=0.5/\delta z, 0.25/\delta z, 0.125/\delta z$ which correspond to correctly
capturing Landau damping on scales of $12 \delta z$, $24 \delta z$, $48 \delta z$,
respectively, where $\delta z=L_z/N_z$, $L_z=1$ for all our runs, and $N_z$
is the number of grid points used in the $z$-direction (taken be $27$
and $54$ for low and high resolution calculations,
respectively). Thus, $k_L=0.25/\delta z$ corresponds to correctly capturing
Landau damping for modes with wavelengths comparable to the size of
the box in the low resolution runs.

The term $\nu_{eff}$ in equations (\ref{eq:kappa_par}) \&
(\ref{eq:kappa_perp}) is an effective collision frequency which is
equal to the real collision frequency $\nu$ as long as $\mu$
conservation is satisfied. However, when the pressure anisotropy is large
enough to drive microinstabilities that
break $\mu$ invariance and enhance pitch angle scattering, then there is
an increase in the effective collision frequency that
decreases the
associated conductivities.  The expressions for $\nu_{eff}$ are given
in equations (\ref{eq:nueff1}), (\ref{eq:nueff2}), and
(\ref{eq:nueff3}) of Appendix \ref{app:pthresh}.

Shearing periodic boundary conditions appropriate to the shearing box
are described in \citet{haw95}. Excluding $V_y$, all variables at the inner $x-$
boundary are mapped to sheared ghost zones at the outer boundary; a
similar procedure applies for the inner ghost zones. $V_y$ has a jump of
$(3/2) \Omega L_x$ across the box while applying the $x-$ shearing
boundary conditions, to account for the background shear in $V_y$.

\subsection{Numerical Methods}

We have used a version of the ZEUS code modified to include kinetic
effects \citep[see][]{sto92a,sto92b}.  ZEUS is a time explicit,
operator split, finite difference algorithm on a staggered mesh, i.e.,
scalars and the diagonal components of second rank tensors are zone
centered, while vectors are located at zone faces, and pseudovectors
and offdiagonal components of second rank tensors are located at the
edges. The location of different variables on the grid is described in
more detail in Appendix \ref{app:grid}. Appendix \ref{app:courant}
describes how we choose the time step $\delta t$ to satisfy the Courant
condition (which is modified by pressure anisotropy and heat
conduction). We also require that the choice of $\delta t$ maintain
positivity of $p_\Par$ and $p_\Perp$.

Implementation of the shearing box boundary conditions is described in
\citet{haw95}.  One can either apply boundary conditions on the components of
${\bf B}$ or the EMF.  We apply shearing periodic boundary conditions
on the EMF to preserve the net vertical flux in the box, although
applying boundary conditions directly on ${\bf B}$ also gives
satisfactory results.

Equations (\ref{eq:SB4}) and (\ref{eq:SB5}) are split into a transport
and source step, analogous to the energy equation in the original MHD
formalism. The transport step is advanced conservatively, and source
step uses central differences in space. It should be noted that in
equation (\ref{eq:SB5}) the $\nabla \cdot {\bf q_\Perp}$ term is not
purely diffusive, and it is necessary to carefully treat the magnetic
gradient part of $q_\Perp$ in the transport step for robustness of the
code (Appendix \ref{app:qperp_conservative}).

We have carried out a series of tests of our newly added subroutines
for evolving anisotropic pressure and parallel heat conduction.  We
tested the anisotropic conduction routine by initializing a ``hot''
patch in circular magnetic field lines and assessing the extent to
which heat remains confined along the field.  This is the same
test described in detail in ~\citet{par05} and we find good agreement
with their results.  For the diffusion of a narrow temperature pulse
in 1D, the anisotropic conduction routine also gave results nearly
identical to that predicted analytically by the 1D diffusion
equation.\footnote{For details of the test and error analysis see:
http://w3.pppl.gov/\~{}psharma/cartesian/1Dheatdiffusion/}
Additional tests of the code included linear waves and instabilities
in non-rotating anisotropic plasmas, including the Alfv\'en wave and
the firehose and mirror instabilities.  
For mirror simulations we observe the formation of stationary
anticorrelated  
density and magnetic structures as seen in the hybrid simulations of 
\citet{mck93}. For firehose we see the instability with magnetic perturbations 
developing at small scales but during saturation the perturbations are at
larger scales as seen in \citet{que96}. 

Finally, the numerical growth rates of the 
kinetic MRI were compared to the analytic results 
for different pressure anisotropies,
$(k_x, k_z)$, collision frequencies, and angles between the magnetic
field and ${\bf \hat{z}}$; we find good agreement with the results of
\citet{qua02} and \citet{sha03}. 
{When $k_L = k_\Par$, the growth rate of the fastest
growing mode is within $\sim 3 \%$ of the theoretical prediction.
The case $B_\phi=B_z$ shows $\sim
2$ faster growth as compared to $B_z=0$ as predicted by linear theory.}

\subsection{Shearing Box and Kinetic MHD}

Certain analytic constraints on the properties and energetics of
shearing box simulations have been described in \citet{haw95}. These constraints
serve as a useful check on the numerical simulations. Here we mention
the modifications to these constraints in KMHD.
Conservation of total energy in the shearing box gives 
\be
\label{eq:encon}
\frac{\partial}{\partial t} \Gamma = \frac{3}{2} \Omega L_x \int_x dA 
\left [ \rho V_x \delta V_y - 
\left ( 1- \frac{4\pi(p_\Par-p_\Perp)}{B^2} \right) 
\frac{B_x B_y}{4\pi} \right],
\ee 
where $\delta V_y = V_y + (3/2) \Omega x$, and $\Gamma$ is the total energy given by,\
\be
\label{eq:energy}
\Gamma = \int d^3x \left [ \rho \left( \frac{V^2}{2} + \phi \right) +
\frac{p_\Par}{2} + p_\Perp + \frac{B^2}{8\pi} \right] \ee where
$\phi=-3/2 \Omega^2 x^2$ is the tidal effective potential about
$R_0$.  Equation~(\ref{eq:encon}) states that the change in the total
energy of the shearing box is due to work done on the box by the
boundaries. Notice that there is an anisotropic pressure contribution
to the work done on the box. Equation (29) in \citet{bal98} for
conservation of angular momentum in cylindrical geometry is also
modified because of the anisotropic pressure and is given by 
\ba
\label{eq:angmom}
\nonumber
\frac{\partial}{\partial t} (\rho R V_\phi) &+& 
\nabla \cdot  \Bigg[ \rho V_\phi {\bf V}R - 
\frac{B_\phi}{4\pi}\left(1-\frac{4\pi(p_\Par-p_\Perp)}{B^2} \right)
{\bf B_p}R \\
&+& \left( p_\Perp + \frac{B_p^2}{8\pi} \right){\bf
\hat{\phi}}R \Bigg] = 0, 
\ea 
where ${\bf B_p} = B_R {\bf \hat{R}}+B_z
{\bf \hat{z}}$ is the poloidal field. 
We can calculate the level of angular momentum transport in our
simulations by measuring the stress tensor given by \be W_{xy}= \rho
V_x \delta V_y - \frac{B_x B_y}{4\pi} + \frac{(p_\Par-p_\Perp)}{B^2}
B_x B_y \ee Note that the stress tensor has an additional contribution
due to pressure anisotropy. One can define a dimensionless stress via
Shakura and Sunyaev's $\alpha$ parameter by \be \alpha \equiv 
\frac{W_{xy}}{P_0} = \alpha_{R} + \alpha_{M} + \alpha_{A} \ee where
$\alpha_R$, $\alpha_M$, $\alpha_A$ are the Reynolds, Maxwell and
anisotropic stress parameters, respectively. As in \citet{haw95} we normalize
the stress using the initial pressure to define an $\alpha$ 
parameter. 

\subsection{Shearing Box Parameters and Initial Conditions}
\label{ssec:IC}

The parameters for our baseline case have been chosen to match the fiducial
run Z4 of \citet{haw95}. 
The simulation box has a radial size $L_x=1$, azimuthal size
$L_y=2\pi$, and vertical size $L_z=1$.  The sound speed
$V_s=\sqrt{p/\rho}=L_z \Omega$, so that the vertical size is about a
disk scale height (though it is an unstratified box).  The pressure is
assumed to be isotropic initially, with $p_0 = \rho_0 V_s^2 = 10^{-6}$
and $\rho_0 = 1$.
All of our simulations start with a vertical field with $\beta=8\pi
p_0/B_0^2 =400$. 
The fastest growing MRI mode for this choice of parameters is reasonably
well resolved.  We consider two different numerical resolutions: $27
\times 59 \times 27$, and $54 \times 118 \times 54$.  Perturbations
are introduced as initially uncorrelated velocity fluctuations. These
fluctuations are randomly and uniformly distributed throughout the
box. They have a mean amplitude of $|\delta V| = 10^{-3} V_s$.

\section{Results}

The important parameters for our simulations are listed in
Table~\ref{tab:tab1}.  Each simulation is labeled by $Z$ (for the
initial $B_z$
field), and $l$ and $h$ represent low ($27 \times 59 \times 27$) and
high ($54 \times 118 \times 54$) resolution runs, respectively.  We
also include low and high resolution MHD runs for comparison with the
kinetic calculations (labeled by $ZM$).  Our models for heat
conduction and pressure isotropization have several parameters: $k_L$,
the typical wavenumber for Landau damping used in the heat flux
(eqs. [\ref{eq:qpar1}] \& [\ref{eq:qperp1}]), and $\xi$, the parameter
that forces the pressure anisotropy to be limited by $p_\Perp/p_\Par -1 <
2 \xi/\beta_\Perp$ (representing pitch angle scattering due to small
scale mirror modes; eq. [\ref{eq:pitch2}]).  All of our calculations
except $Zl8$, $Zl1$, and $Zh1$ also use the ion cyclotron scattering
``hard wall'' from equation (\ref{eq:pitch3}).  In addition to these
model parameters, Table~\ref{tab:tab1} also lists the results of the
simulations, including the volume and time averaged magnetic and
kinetic energies, and Maxwell, Reynolds, and anisotropic stresses.  As
Table ~\ref{tab:tab1} indicates, the results of our simulations depend
quantitatively -- though generally not qualitatively -- on the
microphysics associated with heat conduction and pressure
isotropization.  Throughout this section we use single brackets
$\langle f \rangle$ to denote a volume average of quantity $f$; we use
double brackets $\langle \langle f \rangle \rangle$ to denote a volume
and time average in the saturated turbulent state, from orbit 5
onwards.
\subsection{Fiducial Run}
\label{sec:fiducial}
We have selected run $Zl4$ as our fiducial model to describe in
detail.  This model includes isotropization by ion cyclotron
instabilities and mirror modes, with the former dominating 
(for $\xi = 3.5$; see \S \ref{sec:mirror_scattering}) except at
early times.  The conductivity is determined by $k_L = 0.5/\delta z$ which
implies that modes with wavelengths $\sim 12 \delta z \sim L_z/2$ are damped
at a rate consistent with linear theory.

\begin{figure} 
\epsscale{.80}
\plotone{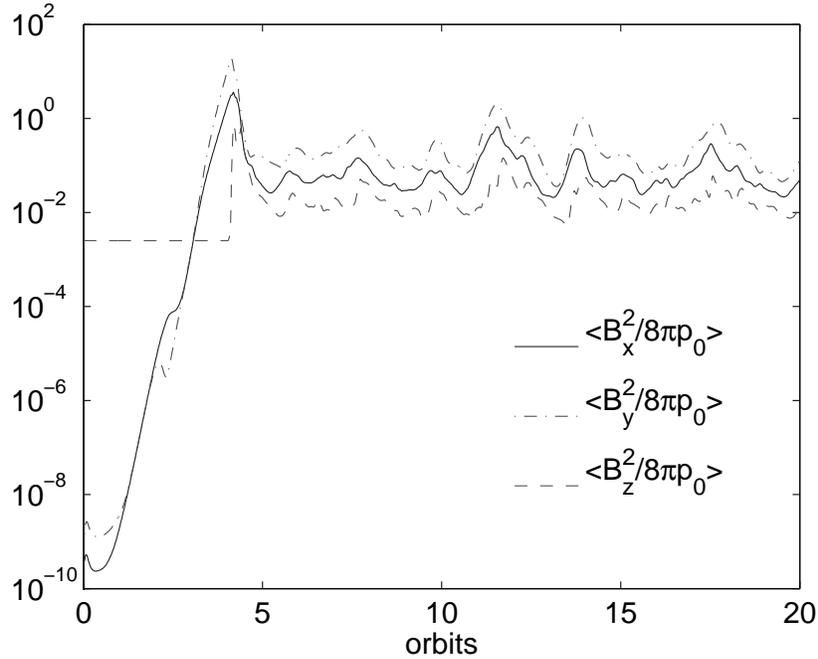}
\caption{Time evolution of volume-averaged magnetic energy for the
fiducial run $Zl4$.  Time is given in number of orbits. There is a
small decrease in the magnetic energy at $\approx 2$ orbits when the
pressure anisotropy is sufficient to stabilize the fastest growing
mode. However, 
small-scale kinetic instabilities limit the magnitude of the pressure
anisotropy, allowing the magnetic field to continue to amplify.
As in MHD, there is a channel phase which breaks down into turbulence
at $\approx 4$ orbits.
\label{fig:figure2}}
\end{figure}

\begin{figure}
\epsscale{.80}
\plotone{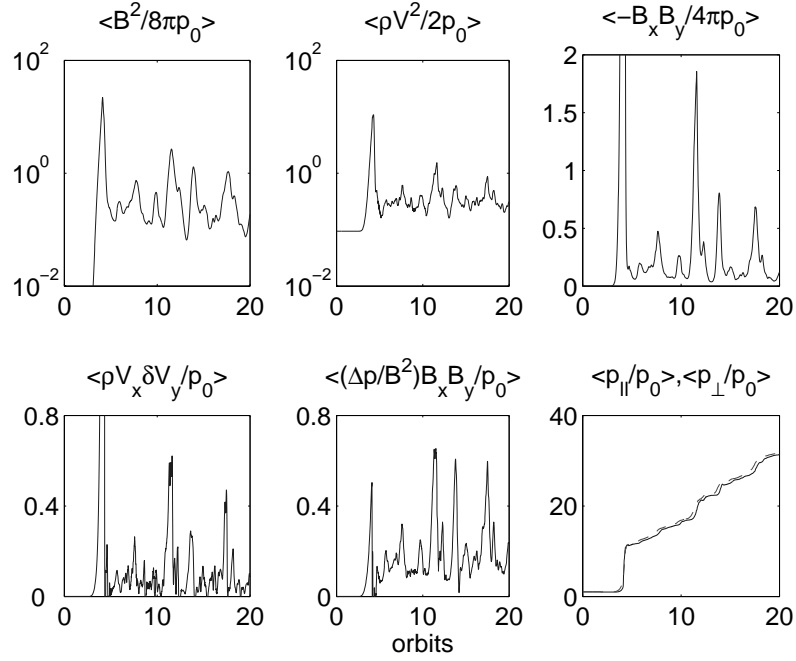}
\caption{Time evolution of volume-averaged magnetic and kinetic
energies, Maxwell, Reynolds, and anisotropic stress, and pressure
($p_\Par$: solid line, $p_\Perp$: dashed line) for the fiducial
model $Zl4$.  Time is given in orbits and
all quantities are normalized to the initial pressure $p_0$. $\delta
V_y=V_y+(3/2)\Omega x $ and $\Delta p = (p_\Par - p_\Perp)$.
\label{fig:figure3}}
\end{figure}

\begin{figure}
\epsscale{.80}
\plotone{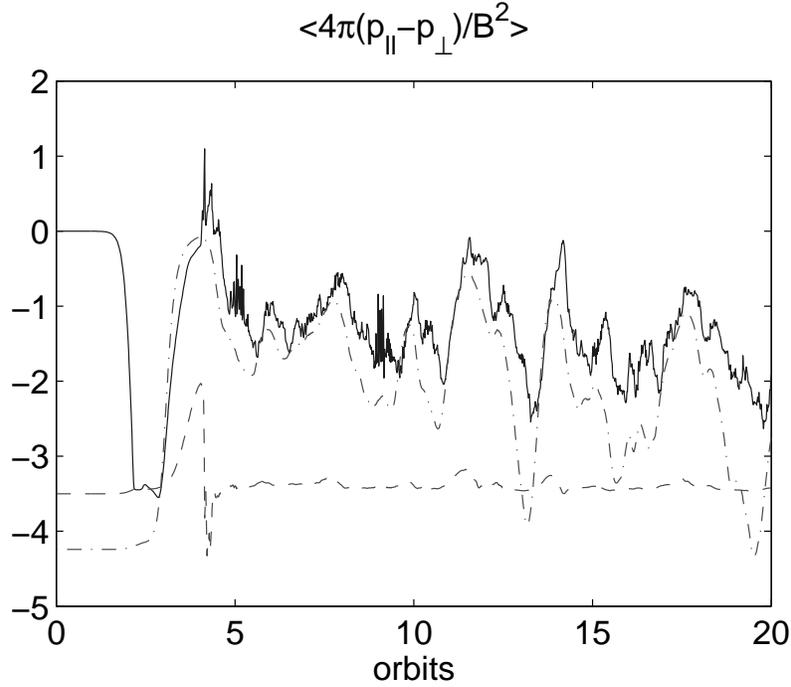}
\caption{Time evolution of volume-averaged pressure anisotropy
($4\pi(p_\Par-p_\Perp)/B^2$: solid line) for model $Zl4$. Also plotted
are the ``hard wall'' limits on the pressure anisotropy due to the ion
cyclotron (dot dashed line) and mirror instabilities (dashed
line). Ion cyclotron scattering is generally more efficient in the
steady state. The limits on pressure anisotropy are applied at each
grid point while this figure is based on volume averaged quantities.
\label{fig:figure4}}
\end{figure}

Figures \ref{fig:figure2}-\ref{fig:figure4} show the time evolution
of various physical quantities for run $Zl4$.  The early linear
development of the instability is similar to that in MHD, with the
field growing exponentially in time. The key new feature is the
simultaneous exponential growth of pressure anisotropy ($p_\Perp >
p_\Par$) as a result of $\mu$ conservation (up to 2 orbits in
Fig. \ref{fig:figure4}).  As described in \S \ref{sec:linmodes},
this pressure anisotropy tends to {\it stabilize} the MRI modes and
shut off the growth of the magnetic field.  Indeed, in simulations
that do not include any isotropization of the pressure tensor, we find
that all MRI modes in the box are stabilized by the pressure
anisotropy and the simulation saturates with the box filled with small
amplitude anisotropic Alfv\'en waves (see Fig.~\ref{fig:figure6}).
This highlights the fact that, unlike in MHD, the MRI is not an exact
nonlinear solution in kinetic theory.  However, the pressure 
anisotropy required to stabilize all MRI modes exceeds the pressure 
anisotropy at which pitch angle scattering due to mirror and ion
cyclotron instabilities become important.  This takes place at about
orbit 2 in run $Zl4$ (see the small `dip' in the growth of B in
Fig. \ref{fig:figure2}), at which point the pressure anisotropy is
significantly reduced and the magnetic field is able to grow to 
nonlinear amplitudes.

The nonlinear saturation at orbit $\sim 5$ appears qualitatively
similar to that in MHD, and may occur via analogues of the parasitic
instabilities described by~\citet{good94}.  The channel solution is,
however, much more extreme in KMHD than MHD (the maximum $B^2$ in
Fig. \ref{fig:figure2} is approximately an order of magnitude larger
than in analogous MHD runs).  
After saturation, the magnetic and
kinetic energies in the saturated state are comparable in KMHD and MHD
(see Table~\ref{tab:tab1}).  This is essentially because the pitch
angle scattering induced by the kinetic microinstabilities acts to
isotropize the pressure, enforcing a degree of MHD-like dynamics on
the collisionless plasma.

Figure \ref{fig:figure3} and Table \ref{tab:tab1} show the various
contributions to the total stress. As in MHD, the Reynolds stress is
significantly smaller than the Maxwell stress.  In kinetic theory,
however, there is an additional component to the stress due to the
anisotropic pressure (eq. [\ref{eq:angmom}]).  In the saturated state,
we find that the Maxwell stress is similar in KMHD and MHD, but that
the anisotropic stress itself is comparable to the Maxwell
stress. Expressed in terms of an $\alpha$ normalized to the initial
pressure, our fiducial run $Zl4$ has $\alpha_M=0.23$,
$\alpha_R=0.097$, and $\alpha_A=0.2$, indicating that stress due to
pressure anisotropy is dynamically important.

Nearly all physical quantities in
Figures~\ref{fig:figure2}-\ref{fig:figure4} reach an approximate
statistical steady state.  The exceptions are that $p_\Par$ and
$p_\Perp$ increase steadily in time because the momentum flux on the
boundaries does work on the system (Eq.\ 45), which is eventually
converted to heat in the plasma by artificial viscosity
and there is no cooling (the same is true in HGB's MHD
simulations).
Because of the steadily increasing internal energy and approximately
fixed $B^2$ (although with large fluctuations), the plasma $\beta$ shows 
a small secular increase from
orbits 5-20 (a factor of $\approx 3$ increase, though with very
large fluctuations due to the large fluctuations in magnetic energy).
Figure \ref{fig:figure4} shows the pressure anisotropy thresholds due
to the ion cyclotron and mirror instabilities, in addition to the
volume averaged pressure anisotropy in run $Zl4$.  From equation
(\ref{eq:pitch3}), the ion cyclotron threshold is expected to scale as
$\sqrt{\beta_\Par}$, which is reasonably consistent with the trend in
Figure \ref{fig:figure4}.  The actual pressure anisotropy in the
simulation shows a small increase in time as well, although less than
that of the ion cyclotron threshold.  These secular changes in $\beta$
and $\Delta p$ are a consequence of the increasing internal energy in
the shearing box, and are probably not realistic.  In a global disk, we
expect that -- except perhaps near the inner and outer boundaries --
$\beta$ will not undergo significant secular changes in time.  
In a small region of a real disk in statistical equilibrium, the heating
would be balanced by radiation or by cooler plasma entering at large $R$
and hotter plasma leaving at small $R$.

\begin{figure}
\epsscale{.80}
\plotone{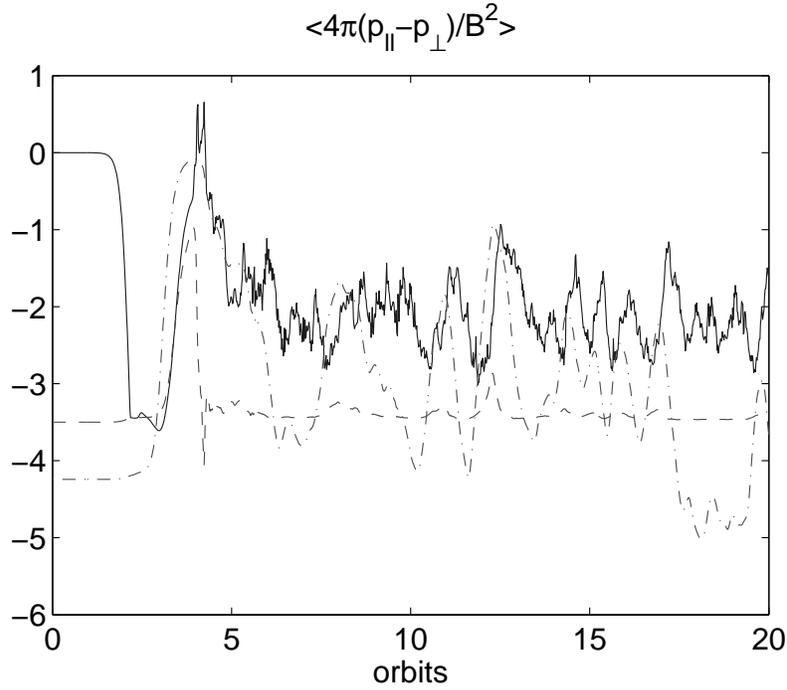}
\caption{Time evolution of volume-averaged pressure anisotropy
($4\pi(p_\Par-p_\Perp)/B^2$: solid line) for model $Zl8$. Also plotted
are the ``hard wall'' limits on the pressure anisotropy due to the ion
cyclotron (dot dashed line) and mirror instabilities (dashed line),
although the ion cyclotron scattering limit is {\it not} applied in
this simulation. The volme averaged pressure anisotropy saturates at
smaller anisotropy than
the mirror threshold at $\xi=3.5$, which is the only limit on pressure 
anisotropy used.
\label{fig:figure5}}
\end{figure}

\begin{figure}
\epsscale{.80}
\plotone{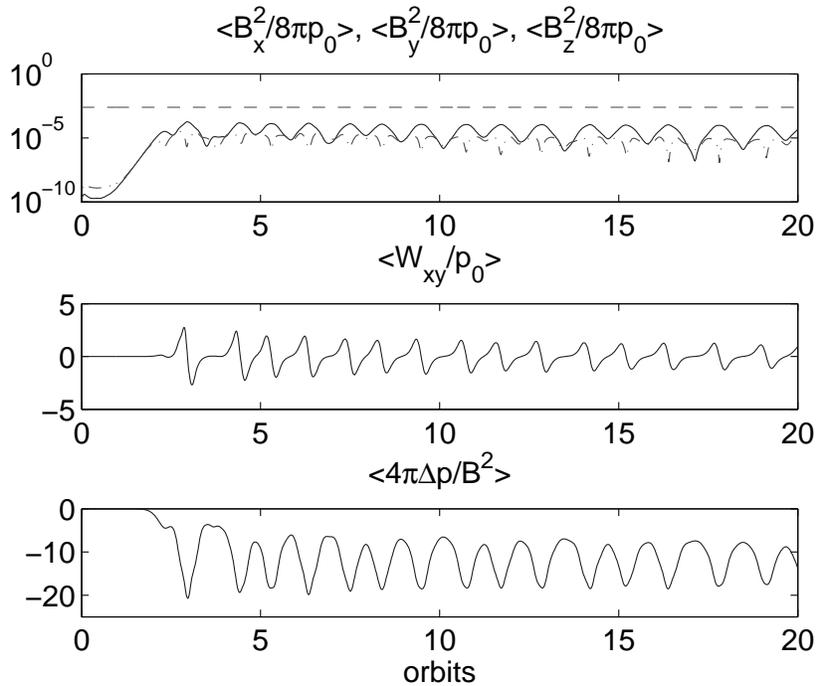}
\caption{Time evolution of volume-averaged magnetic energy (dashed
line: $B_z^2/8\pi p_0$, solid line: $B_x^2/8\pi p_0$, dot dashed line:
$B_y^2/8\pi p_0$), total stress ($W_{xy}/p_0$) in units of $10^{-3}$,
and pressure anisotropy for model $Zl1$. Time is given in orbits and
all quantities are normalized to the initial pressure $p_0$. $\delta
V_y=V_y+(3/2)\Omega x$ and $\Delta p = (p_\Par - p_\Perp)$.  In this
calculation there is no heat conduction and no isotropization of the
pressure tensor.  All resolved MRI modes are thus stabilized by
pressure anisotropy and the `saturated' state is linear anisotropic
Alfv\'en waves with no net angular momentum transport.
\label{fig:figure6}}
\end{figure}

It is interesting to note that in Figure \ref{fig:figure4}, the
pressure anisotropy ($4 \pi \Delta p/B^2$) is closely tied to the ion
cyclotron threshold at times when $B^2$ is rising (which corresponds
to the channel solution reemerging).  Increasing $B$ leads to a
pressure anisotropy with $p_\Perp>p_\Par$ by $\mu$ conservation.
At the same time, the ion cyclotron threshold ($\sim \sqrt{\beta}$)
decreases and thus the threshold is encountered which limits the
anisotropy.  When $B$ is decreasing, however, we do not find the same
tight relationship between the pressure anisotropy and the imposed
threshold.  Figure \ref{fig:figure4} clearly indicates that in our
fiducial simulation pitch angle scattering is dominated by the ion
cyclotron threshold.  For comparison, Figure~\ref{fig:figure5}
shows the pressure anisotropy and thresholds for run $Zl8$ which is
identical to the fiducial run except that the ion cyclotron threshold
is not used and the only scattering is due to the mirror threshold. In
this case, the saturated pressure anisotropy is somewhat larger than
in the fiducial run, but the pressure anisotropy is {\it not} tied to
the mirror threshold. 

Table~\ref{tab:tab2} gives the mean, standard deviation, and standard
error in the mean, for various
quantities in the saturated portion of the fiducial simulation.  The
standard errors are estimated by taking into account the finite correlation
time for the physical quantities in the simulation, as described in
Appendix ~\ref{app:error_bars}.  In
many cases, the deviations are significantly larger than the mean.  As
in MHD, we find that the magnetic energy is dominated by the $y-$component,
which is about a factor of $3$ larger than the $x-$component; the vertical
component is smaller yet. The radial and
azimuthal kinetic energy fluctuations are comparable, while the
vertical component is smaller.
We also find that, as in MHD, the perturbed kinetic and magnetic
energies are not in exact equipartition: the magnetic energy is consistently
larger.
Table~\ref{tab:tab2} also shows the mean and deviations for $\langle
p_\Perp/B \rangle$ and $\langle p_\Par B^2/\rho^2 \rangle$.  Because
of pitch angle scattering $\mu= \langle p_\Perp/B \rangle$ is no
longer conserved. $\langle p_\Par B^2/\rho^2 \rangle$ varies both
because of heat conduction and pitch angle scattering.

The pressure anisotropy in our fiducial run saturates at $4\pi
(p_\Perp-p_\Par)/B^2 \approx 1.5$.  By contrast, the threshold for the
mirror instability is $4 \pi (p_\Perp-p_\Par)/B^2=0.5$. 
This implies
that the model is unstable to generating mirror modes.  However, the
mirror modes that can be excited at this level of anisotropy do not
violate $\mu$ conservation and thus do not contribute to pitch angle
scattering (\S \ref{sec:mirror_scattering}).  They can in principle
isotropize the plasma in a volume averaged sense by spatially
redistributing plasma into magnetic wells \citep[e.g.,][]{kiv96}.  
This saturation mechanism can be calculated using our
kinetic-MHD code and was in fact one of our test problems (for a
uniform plasma).  It does not appear to be fully efficient in the
saturated state of our turbulent disk simulations, even at the highest
resolutions we have run.

In the next few sections we compare the fiducial simulation described
above with variations in the pitch angle scattering model and the
parallel conductivity.  A comparison of the total stress in all of our
simulations is shown in Figure~\ref{fig:figure7}.

\subsection{The Double Adiabatic Limit}
\label{sec:DA}

Simulations $Zl1$ and $Zh1$ are simulations in the double adiabatic
limit (no heat conduction), with no limit on the pressure anisotropy
imposed. In this limit both $\mu= \langle p_\Perp/B \rangle$ and
$\langle p_\Par B^2/\rho^2 \rangle$ are
conserved. Figure~\ref{fig:figure6} shows volume averages of various
quantities as a function of time for the run $Zl1$.  These
calculations are very different from the rest of our results and show
saturation at very low amplitudes ($\delta B^2/B^2 \approx 0.04$). In
the saturated state, the box is filled with shear modified anisotropic
Alfv\'en waves and all physical quantities are oscillating in
time. The total stress is also oscillatory with a vanishing mean,
resulting in negligible transport.  In these calculations, the
pressure anisotropy grows to such a large value that it shuts off the
growth of all of the resolved MRI modes in the box.
Table~\ref{tab:tab1} shows that $\langle \langle 4 \pi
(p_\Par-p_\Perp)/B^2 \rangle \rangle$ saturates at $-11.96$ and
$-10.2$ for the low and high resolution runs, respectively (although
the normalized pressure anisotropy $\langle \langle
(p_\Par-p_\Perp)/p_\Par \rangle \rangle \approx -0.07$ is quite
small).  This is much larger than the anisotropy thresholds for pitch
angle scattering described in \S2.  As a result, we do not expect
these cases to be representative of the actual physics of
collisionless disks.  These cases are of interest, however, in
supporting the predictions of the linear theory with anisotropic
initial conditions considered in $\S$~\ref{sec:linmodes}, and in
providing a simple test for the simulations.  They also highlight the
central role of pressure isotropization in collisionless dynamos.

\subsection{Varying Conductivity}

We have carried out a series of simulations with different
conductivities defined by the parameter $k_L$.  Simulations $Zl2$ and
$Zh2$ are in the CGL limit with vanishing parallel heat conduction,
but with the same limits on pressure anisotropy as the fiducial model.
Simulations $Z6$ use $k_L \delta z = 0.25$ while run $Zl7$ uses
$k_L=0.125/\delta z$.  Both of these are smaller than the value of $k_L \delta 
z = 0.5$ in the fiducial run, which implies a larger conductivity.
Figure~\ref{fig:figure7} shows that the total stress varies by about a
factor of 2 depending on the conductivity and resolution.  Simulations
with larger conductivity tend to have smaller saturation amplitudes
and stresses.  This could be because larger conductivity implies more
rapid Landau damping of slow and fast magnetosonic waves.  In all
cases, however, the anisotropic stress is comparable to the Maxwell
stress as in the fiducial run.  Until a more accurate evaluation is
available of the heat fluxes for modes of all wavelengths in the
simulation simultaneously (either by  a more complete evaluation of the
nonlocal heat fluxes, eqs. [\ref{eq:nonlocal1}-\ref{eq:nonlocal2}], or
even by a fully kinetic MHD code that directly solves eq. [\ref{eq:DKE}],
it is difficult to ascertain
which value of the conductivity best reflects the true physics of
collisionless disks.

\subsection{Different Pitch Angle Scattering Models}
\label{sec:pitchangle}

In this section we consider variations in our model for pitch angle
scattering by high frequency waves.  All of these calculations utilize
$k_L = 0.5/\delta z$.  We note again that the appropriate pitch angle 
scattering model remains somewhat uncertain, primarily because of
uncertainties in the nonlinear saturation of long-wavelength
$\mu-$conserving mirror modes.  The calculations reported here cover
what we believe is a plausible range of models.

Models $Zl5$ and $Zh5$ place a more stringent limit on the allowed
pressure anisotropy, taking $\xi=0.5$ in equation
(\ref{eq:pitch2}).  This corresponds to the threshold of the mirror
instability.  Not surprisingly, this simulation is the most
``MHD-like'' of our calculations, with magnetic and kinetic energies
and Maxwell stresses that are quite similar to those in MHD.  Even
with this stringent limit, however, the anisotropic stress is $\approx
1/3$ of the Maxwell stress.  It is also interesting to note that
although the dimensionless pressure anisotropy is quite small $\langle
\langle 4 \pi (p_\Par-p_\Perp)/B^2 \rangle \rangle \approx -0.02$, the
dimensionless anisotropic stress $\langle \langle 4 \pi
(p_\Par-p_\Perp)/B^2 \times B_xB_y/p_0 \rangle \rangle \approx -0.07$
is significantly larger (and larger than Reynolds stress) because of
correlations between the pressure anisotropy and field strength.

As a test of how large a collisionality is needed for the results of
our kinetic simulations to rigorously approach the MHD limit, we have
carried out a series of simulations including an explicit
collisionality $\nu$ and varying its magnitude relative to the disk
frequency $\Omega$. Our results are summarized in Table
\ref{tab:tab3} and Figure~\ref{fig:figure8}.  
In these simulations we start with initial conditions
determined by the saturated turbulent state of our fiducial run $Zl4$,
but with an explicit collision frequency (in addition to the
scattering models described in \S \ref{sec:anisotropy}).  Figure
\ref{fig:figure8} shows that for $\nu/\Omega \lesssim 20$, the results
are very similar to the collisionless limit.  For larger collision
frequencies the anisotropic stress is reduced and the simulations
quantitatively approach the MHD limit.  These results are similar to
those obtained by \citet{sha03}, who 
found that in linear 
calculations the MHD limit for modes with $k \sim \Omega/v_A$
is approached when $\nu \gtrsim \sqrt{\beta}\,\Omega$.

To consider the opposite limit of lower collisionality, run $Zl8$
places a less stringent limit on the allowed pressure anisotropy,
taking $\xi=3.5$ in equation (\ref{eq:pitch2}) and ignoring the limit
set by the ion cyclotron instability in equation (\ref{eq:pitch3}).
The results of this calculation are not physical but are useful for
further clarifying the relative importance of the Maxwell and
anisotropic stresses as a function of the pitch angle scattering
rate.  In $Zl8$, the saturated magnetic energy and Maxwell stress are
lower than in all of our other calculations (excluding the double
adiabatic models described in \S\ref{sec:DA}).  Interestingly,
however, the total stress is comparable to that in the other
calculations (Fig. \ref{fig:figure7}) because the anisotropic stress
is $\approx 2.4$ times larger than the Maxwell stress (Table 1).
As discussed briefly in $\S$ \ref{sec:fiducial}, the pressure
anisotropy in this simulation is not simply set by the applied mirror
pitch angle scattering threshold (see Fig. \ref{fig:figure5}).  It
is possible that resolved mirror modes contribute to decreasing the
volume averaged pressure anisotropy (but see below).

Finally, in models $Z3$ we include parallel heat conduction but do not
limit the pressure anisotropy.  In these calculations, we expect to be
able to resolve the long-wavelength $\mu$-conserving mirror modes that
reduce the pressure anisotropy by forming magnetic wells \citep[as
in][]{kiv96}.\footnote{At the resolution of $Zl3$, the fastest growing
mirror mode in the computational domain has a linear growth comparable
to that of the MRI.}  In our test problems with uniform anisotropic
plasmas, this is precisely what we find.  In the shearing box
calculations, however, even at the highest resolutions, we find that
the pressure anisotropy becomes so large that equations
(\ref{eq:pitch2}) and (\ref{eq:pitch3}) are violated and pitch angle
scattering due to high frequency microinstabilities would become
important.  The resolved mirror modes are thus not able to isotropize
the pressure sufficiently fast at all places in the box.\footnote{In
higher resolution simulations, one can resolve smaller-scale and
faster growing mirror modes, and thus the effects of isotropization by
resolved mirror modes could be come increasingly important.  We see no
such indications, however, for the range of resolutions we have been
able to simulate.}
However, it is hard to draw any firm conclusions
from these simulations because they stop at around 4 orbits (for both
resolutions $Zl3$ and $Zh3$) during the initial nonlinear transient
stage.  At this time the pressure becomes highly anisotropic and 
becomes very small at some grid points, and the 
time step limit causes $\delta t \rightarrow 0$.

\section{Summary \& Discussion}

In this paper we have carried out local shearing box simulations of
the magnetorotational instability in a collisionless plasma.  We are
motivated by the application to hot radiatively inefficient flows
which are believed to be present in many low-luminosity accreting
systems.  Our method for simulating the dynamics of a collisionless
plasma is fluid-based and relies on evolving a pressure tensor with
closure models for the heat flux along magnetic field lines (\S2).
These heat flux models can also be thought of as approximating the
collisionless (Landau) damping of linear modes in the simulation.

By adiabatic invariance, a slow increase (decrease) in the magnetic
field strength tends to give rise to a pressure anisotropy with
$p_\Perp > p_\Par$ ($p_\Par > p_\Perp$), where the directions are
defined by the local magnetic field.  Such a pressure anisotropy can,
however, give rise to small scale kinetic instabilities (firehose,
mirror, and ion cyclotron) which act to isotropize the pressure
tensor, effectively providing an enhanced rate of pitch angle
scattering (``collisions'').  We have included the effects of this
isotropization via a subgrid model which restricts the allowed 
magnitude of the pressure anisotropy (\S \ref{sec:anisotropy}).

We find that the nonlinear evolution of the MRI in a collisionless plasma
is qualitatively similar to that in MHD, with comparable saturation
magnetic field strengths and magnetic stresses.  The primary new effect in
kinetic theory is the existence of angular momentum transport due to the
anisotropic pressure stress (eq. [\ref{eq:angmom}]).  For the allowed
pressure anisotropies estimated in \S \ref{sec:anisotropy}, the anisotropic
stress is dynamically important and is as large as the Maxwell stress
(Table 1).  The precise rate of transport in the present simulations 
is difficult to quantify accurately and depends
-- at the factor of $\sim 2$ level -- on some of the uncertain microphysics
in our kinetic analysis (e.g., the rate of heat conduction along magnetic
field lines and the exact threshold for pitch angle scattering by
small-scale instabilities; see Fig. \ref{fig:figure7}).  For better
results, it would be interesting to extend these calculations with a more
accurate evaluation of the actual nonlocal heat fluxes,
equations (\ref{eq:nonlocal1})-(\ref{eq:nonlocal2}), or even to directly solve
equation (\ref{eq:DKE}) for the particle distribution function.
Further kinetic studies in the local shearing box, including studies of the
small-scale instabilities that limit pressure anisotropy, would be helpful
in developing appropriate fluid closures for global simulations.

It is interesting to note that two-temperature RIAFs can only be
maintained below a critical luminosity $\sim \alpha^2 L_{\rm EDD}$
\citep[][]{ree82}.  Thus enhanced transport in kinetic theory due to
the anisotropic pressure stress would extend upward in luminosity the
range of systems to which RIAFs could be applicable.  This is
important for understanding, e.g., state transitions in X-ray binaries
\citep[e.g.,][]{esi97}.

In addition to angular momentum transport by anisotropic pressure
stresses, Landau damping of long-wavelength modes can be dynamically
important in collisionless accretion flows.  Because the ZEUS code we
employ is non-conservative, we cannot carry out a rigorous calculation
of heating by different mechanisms such as Landau damping and
reconnection.  Following the total energy-conserving scheme of
~\citet{tur03}, however, we estimate that the energy dissipated by
Landau damping is comparable to or larger than that due reconnection
(which is the major source of heating in MHD simulations).  One caveat
to this analysis is that in local simulations, the pressure increases
in time due to heating, while $B^2 \sim {\rm constant}$.  Thus $\beta$
increases in time and the turbulence becomes more and more
incompressible.  This will artificially decrease the importance of
compressible channels of heating.  Clearly it is of significant interest
to better understand heating and energy dissipation in RIAFs,
particularly for the electrons. We will carry out a more a systematic
analysis of the energetics of collisionless disks in future global
simulations.

In all of our calculations, we have assumed that the dominant source
of pitch angle scattering is high frequency microinstabilities
generated during the growth and nonlinear evolution of the MRI. We
cannot, however, rule out that there are other sources of high
frequency waves that pitch angle scatter and effectively decrease the
mean free path of particles relative to that calculated here (e.g.,
shocks and reconnection).  As shown in 
Table \ref{tab:tab3} and Figure \ref{fig:figure8}, this would decrease 
the magnitude of the anisotropic stress;
we find that for $\nu \gtrsim 30 \, \Omega$, the results of our
kinetic simulations quantitatively approach the MHD limit. In this
context it is important to note that the incompressible part of the MHD
cascade launched by the MRI is expected to be highly anisotropic with
$k_\Perp \gg k_\Par$ \citep{gol95}.  As a result, there is very little
power in high frequency waves that could break $\mu$ conservation.
It is also interesting to note that
satellites have observed that the pressure anisotropy in the solar
wind near 1 AU is approximately marginally stable to the firehose
instability \citep{Kasper02}, consistent with our assumption that 
microinstabilities dominate the isotropization of the plasma.

In this paper we have focused on kinetic modifications to angular
momentum transport via anisotropic pressure stresses and parallel heat
conduction.  In addition,
kinetic effects substantially modify the stability of thermally
stratified low collisionality plasmas such as those expected in RIAFs.
\citet{bal00} showed that in the presence of anisotropic heat
conduction, thermally stratified plasmas are unstable when the {\it
temperature} decreases outwards, rather than when the entropy decreases
outwards (the usual Schwarzschild criterion).  This has been called the
magnetothermal instability (MTI).
\citet{par05} show that
in nonrotating atmospheres the MTI leads to magnetic field
amplification and efficient heat transport.  In future global
simulations of RIAFs, it will be interesting to explore the combined
dynamics of the MTI, the MRI, and angular momentum transport via
anisotropic pressure stresses.

\acknowledgments

We are grateful to Stephane Ethier for his help with MPI. We thank
Shigenobu Hirose for sharing his parallel shearing box version of ZEUS
with us. We are grateful to Scott Klasky for his help on
visualization.  Finally, we thank Ian Parrish for his anisotropic
conduction test, and doing comparisons with us.  EQ is supported in
part by NSF grant AST 0206006, NASA grant NAG5-12043, an Alfred
P. Sloan Fellowship, and the David and Lucile Packard Foundation.
Part of this research was carried out during a sabbatical by GWH at the
University of California at Berkeley, where he is thankful for the
hospitality and support of the Miller Institute for Basic Research in
Science.  This work was also done in part at the Kavli Insitute for
Theoretical Physics, supported by the National Science Foundation
under Grant No. PHY99-0794.

\appendix

\section{Appendix}
\subsection{Grid and Variables}
\label{app:grid}

Figure \ref{fig:figure9} shows the location of variables on the
grid. Scalars and diagonal components of second rank tensors ($\rho$,
$p_\Par$, and $p_\Perp$) are zone centered. Vectors, representing
fluxes out of the box, are located at the cell faces (${\bf V}$, ${\bf
B}$, and ${\bf q_{\Par, \Perp}}$). The inductive electric field ({\bf
E}) is located at cell edges such that the contribution of each edge
in calculating $\oint {\bf E} \cdot {\bf dl}$ over the whole box
cancels, and $\nabla \cdot {\bf B}=0$ is satisfied to machine
precision. The off diagonal part of the pressure tensor in Cartesian
coordinates is related to
${\bf \Pi} = {\bf \hat{b}\hat{b}}(p_\Par-p_\Perp)$. This is a
symmetric tensor whose components $P{xy}$, $P{xz}$, and $P{yz}$ are
located such that the finite difference formulae for the evolution of
velocities due to off diagonal components of stress are given by \ba
&& {Vx_{i,j,k}}^{n+1}={Vx_{i,j,k}}^n -
\frac{\delta t}{\delta y}(P{xy}_{i,j+1,k}^n-P{xy}_{i,j,k}^n) -
\frac{\delta t}{\delta z}(P{xz}^n_{i,j,k+1}-P{xz}_{i,j,k}^n), \\ &&
{Vy_{i,j,k}}^{n+1}={Vy_{i,j,k}}^n -
\frac{\delta t}{\delta x}(P{xy}_{i+1,j,k}^n-P{xy}_{i,j,k}^n) -
\frac{\delta t}{\delta z}(P{yz}^n_{i,j,k+1}-P{yz}_{i,j,k}^n), \\ &&
{Vz_{i,j,k}}^{n+1}={Vz_{i,j,k}}^n - 
\frac{\delta t}{\delta x}(P{xz}_{i+1,j,k}^n-P{xz}_{i,j,k}^n) -
\frac{\delta t}{\delta y}(P{yz}^n_{i,j+1,k}-P{yz}_{i,j,k}^n).  \ea 

\subsection{Determination of $\delta t$: stability, and positivity}
\label{app:courant}
A time explicit algorithm must limit the time step in order to satisfy
the Courant-Friedrichs-Levy (CFL) stability condition. Physically,
$\delta t$ must be smaller than the time it takes any signal (via fluid or
wave motion) to cross one grid zone. There is also a limit imposed on
$\delta t$ for numerical stability of the diffusive steps. Additionally,
since there are quantities which must be positive definite ($\rho$,
$p_\Par$, $p_\Perp$), we also require $\delta t$ to satisfy positivity. We
adopt the following procedure to choose $\delta t$: \ba && \delta t_{adv} =
\frac{\mbox{min}\{\delta x,\delta y,\delta z\}}{
(|V|+|V_A|+|V_s|+|\Omega L_x|)}, \\ &&
\delta t_\Par = \frac{\mbox{min}\{\delta x^2,\delta y^2,\delta z^2\}}
{2 \kappa_\Par}, \\
&& \delta t_\Perp = \frac{\mbox{min}\{\delta x^2,\delta y^2,\delta z^2\}}{2 
\kappa_\Perp}, \ea where $V_A=B/\sqrt{4\pi}$ is the Alfv\'en speed,
and $V_s=\mbox{max}\{ \sqrt{3 p_\Par/\rho}, \sqrt{2 p_\Perp/\rho} \}$
is the maximum sound speed, taking the anisotropy into account. 
$\delta t_{adv}$, $\delta t_\Par$, and and $\delta t_\Perp$ correspond
to limits on the time step for stability to advection, and parallel
and perpendicular heat conduction, respectively.

The source steps for $p_\Par$ and $p_\Perp$ are given by \ba &&
\frac{p_\Par^{n+1}-p_\Par^n}{\delta t} = \left ( -\nabla \cdot {\bf q_\Par}
- 2 p_\Par {\bf \hat{b}} \cdot \nabla {\bf V} \cdot {\bf \hat{b}} + 2
q_\Perp \nabla \cdot \hat{b} \right)^n = A1, \\ &&
\frac{p_\Perp^{n+1}-p_\Perp^n}{\delta t} = \left ( -\nabla \cdot {\bf
q_{\Perp T} } - p_\Perp \nabla \cdot {\bf V} + p_\Perp {\bf \hat{b}}
\cdot \nabla {\bf V} \cdot {\bf \hat{b}} - q_\Perp \nabla \cdot
\hat{b} \right)^n = A2, \ea where ${\bf q_{\Perp T} }=-\kappa_\Perp
\nabla_\Par T_\Perp$ denotes the temperature gradient part of ${\bf
q_\Perp}$. For positivity of $p_\Par^{n+1}$ and $p_\Perp^{n+1}$ we
require that the following conditions are satisfied: whenever $A1$ and
$A2$ are negative, $\delta t_{pos} = \mbox{min} \{ -p_\Par^n/A1, -
p_\Perp^n/A2 \}$; if $A1>0$, $A2<0$, then $\delta t_{pos} =
-p_\Perp^n/A2$; if $A1<0$, $A2>0$, then $\delta t_{pos} =
-p_\Par^n/A1$.  Thus, our final constraint on the timestep $\delta t$ is
given by 
\be 
\delta t = C_0 \times \mbox{min} \left \{ 1/[\mbox{max} \{
\delta t_{adv}^{-2} + \delta t_\Par^{-2} + \delta t_\Perp^{-2}
\}]^{1/2}, \mbox{min} \{\delta t_{pos} \} \right \} \ee where the
$\mbox{max}$ and $\mbox{min}$ are taken over all zones in the box and
$C_0$ is a safety factor (Courant Number) which we take to be
$0.5$.

\subsection{Implementation of the Pressure Anisotropy ``hard wall''}
\label{app:pthresh}
If the pressure anisotropy is larger than the constraints given 
in \S2 by equations (\ref{eq:pitch1})-(\ref{eq:pitch3}),  then
microinstabilities will turn on that will enhance the pitch-angle
scattering rate and quickly reduce the pressure anisotropy to near
marginal stability.  
Because this is a numerically stiff problem, we use an implicit 
approach, following the treatment of \citet{bir01}.
Whenever equation~(\ref{eq:pitch1}) is violated,
we use the following prescription for pitch angle scattering:
\ba
\label{eq:fire1}
&& p_\Par^{n+1} = p_\Par^n - \frac{2}{3}\nu_p \delta t \left( \frac{p_\Par^{n+1}}{2}
-p_\Perp^{n+1} - \frac{B^2}{4\pi} \right), \\
\label{eq:fire2}
&& p_\Perp^{n+1} = p_\Perp^n + \frac{1}{3}\nu_p \delta t \left(
\frac{p_\Par^{n+1}}{2} -p_\Perp^{n+1} - \frac{B^2}{4\pi} \right), \ea
where $\nu_p$ is a very large ($\gg 1/\delta t$) rate at which marginal
stability is approached.  This implicit implementation (which can be solved by
inverting a $2 \times 2$ matrix) with large $\nu_p$ ensures that
each time step the pressure anistropy will drop to be very near marginal
stability for the firehose instability to break $\mu$ invariance.
Given this pitch angle scattering, the
collisionality parameter $\nu_{eff}$ in the thermal conductivity (eqs
[\ref{eq:kappa_par}]-[\ref{eq:kappa_mag}]) is obtained by comparing
equations (\ref{eq:fire1}) and (\ref{eq:fire2}) with equations
(\ref{eq:SB4}) and (\ref{eq:SB5}):  \be
\label{eq:nueff1}
\nu_{eff}= \mbox{max} \left\{ \nu_p \frac{\left(
\frac{p_\Par^{n+1}}{2} -p_\Perp^{n+1}-\frac{B^2}{4\pi} \right) }
{\left( p_\Par^{n+1}-p_\Perp^{n+1} \right)}, \nu \right\}.  \ee 
The effective pitch angle scattering rate $\nu_{eff}$ is independent of
$\nu_p$ (and much smaller than $\nu_p$) in the limit of large $\nu_p$, and
is by definition just large enough to balance other terms in equations
(\ref{eq:SB4}-\ref{eq:SB5}) that are trying to
increase the pressure anisotropy beyond marginal stability.

The
prescriptions for pitch angle scattering due to mirror modes and ion
cyclotron waves are similar.  For mirror modes we use \ba &&
p_\Par^{n+1} = p_\Par^n - \frac{2}{3}\nu_p \delta t \left( p_\Par^{n+1}
-p_\Perp^{n+1} + 2 \xi \frac{p_\Par^{n+1}}{\beta_\Perp^{n+1}} \right),
\\ && p_\Perp^{n+1} = p_\Perp^n + \frac{1}{3}\nu_p \delta t \left(
p_\Par^{n+1} -p_\Perp^{n+1} + 2 \xi
\frac{p_\Par^{n+1}}{\beta_\Perp^{n+1}} \right) \ea to limit the
pressure anisotropy ($\xi=3.5$ for our fiducial run $Zl4$) and
$\nu_{eff}$ is given by \be
\label{eq:nueff2}
\nu_{eff}= \mbox{max} \left\{ \nu_p \frac{\left( p_\Par^{n+1}
-p_\Perp^{n+1}+ 2 \xi \frac{p_\Par^{n+1}}{\beta_\Perp^{n+1}} \right) }
{\left( p_\Par^{n+1}-p_\Perp^{n+1} \right)},
\nu \right\}.
\ee
For ion cyclotron pitch angle scattering we use
\ba
&& p_\Par^{n+1} = p_\Par^n - \frac{2}{3}\nu_p \delta t \left( p_\Par^{n+1}
-p_\Perp^{n+1} + S \frac{p_\Par^{n+1}}{\sqrt{\beta_\Par^{n+1}}}  \right), \\
&& p_\Perp^{n+1} = p_\Perp^n + \frac{1}{3}\nu_p \delta t \left( p_\Par^{n+1}
-p_\Perp^{n+1} + S \frac{p_\Par^{n+1}}{\sqrt{\beta_\Par^{n+1}}}  \right),
\ea
and $\nu_{eff}$ is given by
\be
\label{eq:nueff3}
\nu_{eff}= \mbox{max} \left\{ \nu_p \frac{\left( p_\Par^{n+1}
-p_\Perp^{n+1}+ S \frac{p_\Par^{n+1}}{\sqrt{ \beta_\Par^{n+1}} } \right) }
{\left( p_\Par^{n+1}-p_\Perp^{n+1} \right)},
\nu \right\}.
\ee

\subsection{Implementation of the Advective part of $\nabla \cdot {\bf q_\Perp}$}
\label{app:qperp_conservative}
The flux of $p_\Perp$, ${\bf q_\Perp} = q_\Perp {\bf \hat{b}}$, is
given by \be q_\Perp = - \kappa_\Perp \nabla_\Par \left(
\frac{p_\Perp}{\rho} \right) + \left[ \frac{(p_\Par - p_\Perp)}{\rho
\left( \sqrt{\frac{\pi}{2}\frac{p_\Par}{\rho}} k_L + \nu_{eff} \right)
} \frac{{\bf B} \cdot \nabla B}{B^2} \right] p_\Perp = - \kappa_\Perp
\nabla_\Par \left( \frac{p_\Perp}{\rho} \right) + V_{mag} p_\Perp \ee
where the quantity in square brackets can be thought of as an
advection speed due to parallel magnetic gradients.  Because of this
term, ${\bf q_\Perp}$ is not a purely diffusive operator, but also has
an advective part characterized by the velocity $V_{mag}$. If one
treats the advective part via a simple central difference method, it
does not preserve monotonicity.  Instead, to treat the advective part
of ${\bf q_\Perp}$ properly, we include the advective part in the
transport step. After including the advective heat flux in the
transport step, it takes the form \be \frac{\partial p_\Perp}{\partial
t} + \nabla \cdot \left[ ({\bf V} +V_{mag}{\bf \hat{b}} ) p_\Perp
\right] = 0.  \ee Thus, for updating $p_\Perp$ in the transport step
we calculate fluxes on the cell faces using ${\bf V} + V_{mag} {\bf
\hat{b}}$ instead of just ${\bf V}$. The transport step is then
directionally split in the three directions. The procedure for
monotonicity preserving schemes for calculating fluxes is described in
\citet{sto92a}.

\subsection{Error Analysis}
\label{app:error_bars}

\begin{figure}
\epsscale{.80}
\plotone{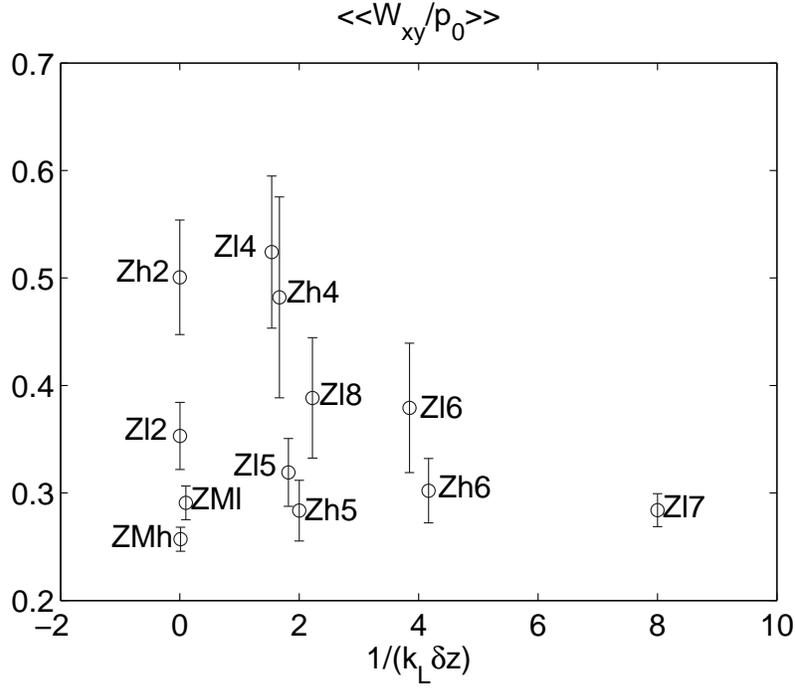}
\caption{Space and time average of the total stress $\langle \langle
W_{xy}/p_0 \rangle \rangle$ versus $1/(k_L \delta z)$ for different
runs.  Error bars shown are based on estimates of the correlation
time of the fluctuations described in \citet{nev05}. 
\label{fig:figure7}} 
\end{figure}

\begin{figure}
\epsscale{.80}
\plotone{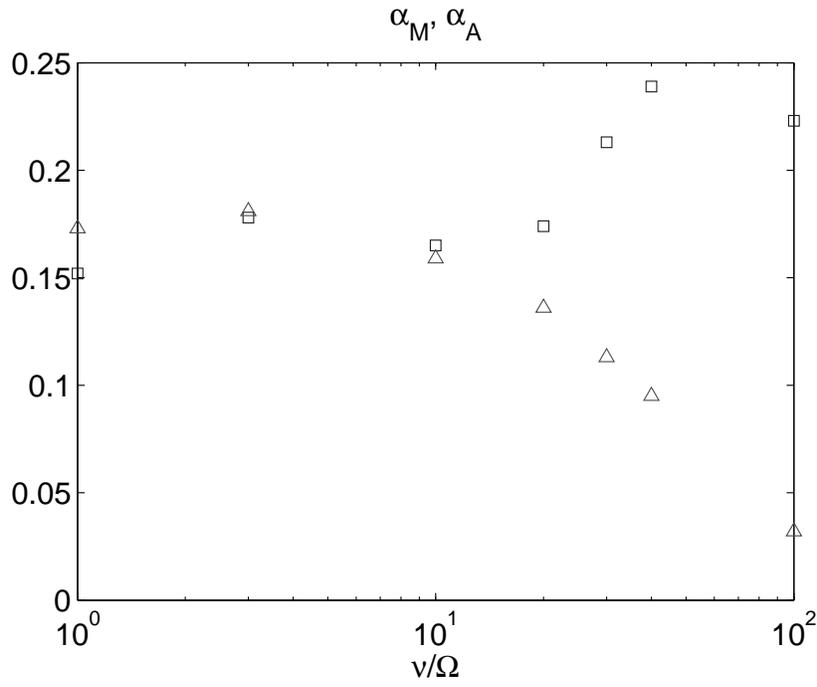}
\caption{Maxwell ($\alpha_M$: squares) and anisotropic stress ($\alpha_A$: triangles)
plotted against the collision frequency normalized to rotation frequency ($\nu/\Omega$).
Transition to MHD occurs for $\nu/\Omega \gtrsim 30$ (see Table~\ref{tab:tab3}).
\label{fig:figure8}}
\end{figure}

\begin{figure}
\psfrag{A}{\large{-($V_x$, $B_x$, $q_{\parallel x}$, $q_{\perp x}$)$_{i,j,k}$}}
\psfrag{B}{\large{$P_{xy_{i,j,k}}$}}
\psfrag{C}{\large{$E_{y_{i,j,k}}$}}
\psfrag{D}{\large{$P_{xz_{i,j,k+1}}$}}
\psfrag{E}{\large{$E_{z_{i,j,k}}$}}
\psfrag{F}{\large{($V_z$, $B_z$, $q_{\parallel z}$, $q_{\perp z}$)$_{i,j,k+1}$}}
\psfrag{G}{\large{-($V_y$, $B_y$, $q_{\parallel y}$, $q_{\perp y}$)$_{i,j,k}$}}
\psfrag{H}{\large{$E_{x_{i,j,k}}$}}
\psfrag{I}{\large{($d$, $p_\parallel$, $p_\perp$)$_{i,j,k}$}}
\psfrag{J}{\large{$P_{yz_{i,j,k+1}} $}}
\psfrag{X}{\large{$x$}}
\psfrag{Y}{\large{$y$}}
\psfrag{Z}{\large{$z$}}
\epsscale{.80}
\plotone{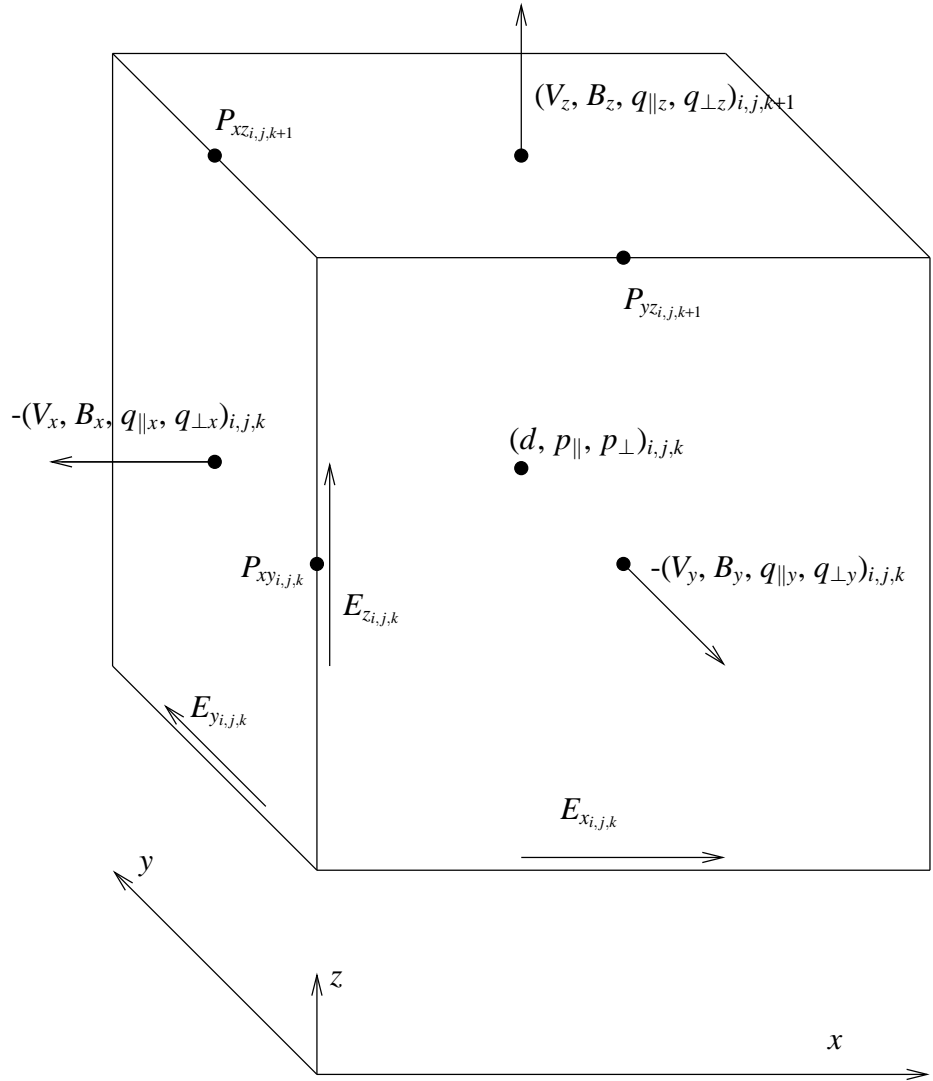}
\caption{ Location of different variables on a 3-D staggered grid.
Vectors ${\bf V}$, ${\bf B}$, and ${\bf q_{\Par,\Perp}}$ are located
at the face centers. Density ($\rho$) and diagonal components of the
pressure tensor ($p_\Perp$, $p_\Par$) are located at the zone
centers. EMF's ($E_x$, $E_y$, $E_z$), and off diagonal components of
the pressure tensor ($P_{xy}$, $P_{xz}$, $P_{yz}$) are located on
appropriate edges.
\label{fig:figure9}}
\end{figure}

The standard errors in the time averages reported in Table~\ref{tab:tab2}
and in Figure~\ref{fig:figure7} are estimated by taking into account the
finite correlation time for the physical quantities in the simulation,
using techniques recommended by \cite{nev05}.  That is, the standard error
for the time average $\langle x \rangle = \int dt \, x(t) / T$ of a signal
$x(t)$ is given by $\sigma_{\langle x \rangle} = \sqrt{{\rm
Var}(x)/N_{eff}}$, where ${\rm Var}(x) = \int dt \, (x(t)-\langle x \rangle
)^2 / T$ is the variance of $x$, $N_{eff} = T/(2 \tau_{int})$ is the
effective number of independent measurements, $T=15$ orbits is the
averaging time, and $\tau_{int}$ is an estimate of the integrated
autocorrelation time.
There are significant subtleties in determining the integrated
autocorrelation time from data.  To deal with this, we use a windowing
technique as recommended by \cite{nev05}, using $\tau_{int} = \int_0^T
d\tau \, C(\tau) W(\tau/\tau_w) $, where $C(\tau)$ is the 2-time
correlation function from the data, $W(\tau/\tau_w)$ is a smooth window
function that cuts off the integral at $\tau \sim \tau_w$, and $\tau_w
\sim \sqrt{T \tau_{int}}$ (this gives results insensitive to the choice of
window width for $\tau_{int} \ll T$).
\cite{Winters03} found from comparing 3 realizations of shearing box MRI
simulations that the magnetic stress had a variation of approximately $\pm
6.5\%$ after averaging over 85 orbits.  The simulations we show here were
averaged over 15 orbits, so extrapolating from \cite{Winters03} one might
expect the uncertainties to be larger by a factor of $\approx \sqrt{85/15}
\approx 2.4$.  This is consistent with the typical error bars we report in
Table~\ref{tab:tab2} and Figure~\ref{fig:figure7}.


\newpage


\clearpage

\begin{deluxetable}{ccrrrrrrrr}
\tabletypesize{\scriptsize}
\tablecaption{Simulation parameters\label{tab:tab1}}
\tablewidth{0pt}
\tablehead{
\colhead{Label} & \colhead{Grid} & \colhead{$k_L^a$} & \colhead{$\xi^b$}
& \colhead{$\langle \langle \frac{B^2}{8\pi p_0} \rangle \rangle^c$} &
\colhead{$\langle \langle \frac{V^2}{2 p_0} \rangle \rangle$} &
\colhead{$\langle \langle \frac{B_x B_y}{4\pi p_0} \rangle \rangle$} &
\colhead{$\langle \langle \frac{\rho V_x \delta V_y}{p_0} \rangle \rangle$} &
\colhead{$\langle \langle \frac{\Delta p^*}{B^2}\frac{B_xB_y}{p_0}
\rangle \rangle$} & 
\colhead{ $ \langle \langle \frac{4\pi \Delta p}{B^2}
\rangle \rangle$}
}
\startdata
$Zl1$ & $27\times 59 \times 27$ & $\infty$ & $\infty$ & $0.0026$ &$0.094$& $0.0$ & $0.0$ & $0.0$ & $-11.96$ \\
$Zl2$ & $27\times 59 \times 27$ & $\infty$ & $3.5$ & $0.25$ & $0.28$ & $0.15$ & $0.067$ &
$0.14$ & $-0.96$ \\
$Zl3^{\dag}$ & $27\times 59 \times 27$ & $0.5/\delta z$ & $\infty$ & $-$ & $-$ & $-$ & $-$ & $-$ & $-$ \\
$Zl4$ & $27\times 59 \times 27$ & $0.5/\delta z$ & $3.5$ & $0.38$ & $0.36$ & $0.23$ & $0.097$ &
 $0.20$ & $-1.37$ \\
$Zl5$ & $27\times 59 \times 27$ & $0.5/\delta z$ & $0.5$ & $0.35$ & $0.27$ & $0.197$ & $0.054$ & $0.069$ & $-0.02$ \\
$Zl6$ & $27\times 59 \times 27$ & $0.25/\delta z$ & $3.5$ & $0.27$ & $0.30$ & $0.16$ & $0.070$ & $0.15$ & $-1.39$ \\
$Zl7$ & $27\times 59 \times 27$ & $0.125/\delta z$ & $3.5$ & $0.21$ & $0.26$ & $0.124$ & $0.051$ & $0.117$ & $-1.44$ \\
$Zl8$ & $27\times 59 \times 27$ & $0.5/\delta z$ & $3.5$ & $0.157$ & $0.315$ & $0.094$ & $0.069$ & $0.225$ & $-2.11$ \\
$ZMl$ & $27\times 59 \times 27$ & $-$ & $-$ & $0.39$ & $0.29$ & $0.22$ & $0.066$ & $-$ & $-$ \\
$Zh1$ & $54\times 118 \times 54$ & $\infty$ & $\infty$ & $0.0026$ & $0.095$ &
$0.0$ & $0.0$ & $0.0$ & $-10.2$ \\
$Zh2$ & $54\times 118 \times 54$ & $\infty$ & $3.5$ & $0.41$ & $0.32$ & $0.24$ & $0.083$
& $0.18$ & $-1.09$ \\
$Zh3^\dag$ & $54\times 118 \times 54$ & $0.5/\delta z$ & $\infty$ &$-$& $-$ & $-$ & $-$ & $-$ & $-$ \\
$Zh4$ & $54\times 118 \times 54$ & $0.5/\delta z$ & $3.5$ & $0.40$ & $0.33$ & $0.22$ & $0.078$ & $0.18$ & $-1.20$ \\
$Zh5$ & $54\times 118 \times 54$ & $0.5/\delta z$ & $0.5$ & $0.349$ & $0.253$ & $0.186$ & $0.042$ & $0.055$ & $-0.02$ \\
$Zh6$ & $54\times 118 \times 54$ & $0.25/\delta z$ & $3.5$ & $0.24$ & $0.26$ & $0.13$ & $0.044$ & $0.13$ & $-1.42$ \\
$ZMh$& $54\times 118 \times 54$ & $-$ & $-$ & $0.375$ & $0.27$ & $0.204$ & $0.0531$ & $-$ & $-$ \\
\enddata

\tablecomments{Vertical field simulation with initial $\beta=400$.
$Z$ indicates that all simulations start with a vertical field, `$l$',
`$h$' indicate low and high resolution runs respectively. $Zl4$ is the
fiducial run. $Zl1$, $Zh1$  are the runs in CGL limit.  $ZMl$ and$ZMh$ are
the MHD runs.}

\tablenotetext{a}{Wavenumber parameter used in Landau closure for
parallel heat conduction (eqs. [\ref{eq:qpar1}] \&
[\ref{eq:qperp1}]).}  \tablenotetext{b}{Imposed limit on pressure
anisotropy for pitch angle scattering due to mirror instability
(eq. [\ref{eq:pitch2}]).  Excluding $Zl1$, $Zh1$, and $Zl8$ all of
these calculations also use a pressure anisotropy limit due to the ion
cyclotron instability (eq. [\ref{eq:pitch3}]).}
\tablenotetext{c}{$\langle \langle \rangle \rangle$ denotes a time and
space average taken from 5 to 20 orbits.}
\tablenotetext{*}{$\Delta p=(p_\Par-p_\Perp)$}
\tablenotetext{\dag}{These cases run for only
$\approx 4$ orbits at which point the time step becomes very small
because regions of large pressure anisotropy are created (see \S
\ref{sec:pitchangle}).}
\end{deluxetable}

\clearpage

\begin{deluxetable}{cccccc}
\tabletypesize{\scriptsize}
\tablecaption{Statistics for Model $Zl4$.\label{tab:tab2}}
\tablewidth{0pt}
\tablehead{
\colhead{Quantity $f$} & \colhead{$\langle \langle f \rangle \rangle$}
& \colhead{$ \langle \langle \delta f^2 \rangle \rangle^{1/2} $}
& \colhead{$(\frac{\tau_{int}}{T})^{1/2} \langle \langle \delta f^2 \rangle 
\rangle^{1/2}$}  & \colhead{$\mbox{min}(f)$} & \colhead{$\mbox{max}(f)$} }
\startdata
$\frac{B_x^2}{8\pi p_0}$ & $0.083$ & $0.092$ & $0.016 $ & $0.021$ & $0.662$\\
$\frac{B_y^2}{8\pi p_0}$ & $0.276$ & $0.318$ & $0.048 $ & $0.036$ & $1.987$\\
$\frac{B_z^2}{8\pi p_0}$ & $0.021$ & $0.017$ & $0.0025 $ & $0.0032$ & $0.144$\\
$\frac{\rho V_x^2}{2 p_0}$ & $0.102$ & $0.094$ & $0.014 $ & $0.0184$ & $0.63$\\
$\frac{\rho \delta V_y^2}{2 p_0}$ & $0.125$ & $0.079$ & $0.0127 $ & $0.715$ & $0.0264$\\
$\frac{\rho V_z^2}{2 p_0}$ & $0.037$ & $0.034$ & $0.0032 $ & $0.008$ & $0.348$\\
$\frac{-B_xB_y}{4\pi p_0}$ & $0.229$ & $0.277$ & $0.0434 $ & $0.037$ & $1.856$\\
$\frac{\rho V_x\delta V_y}{p_0}$ & $0.097$ & $0.113$ & $0.0147 $ & $-0.072$ & $0.6211$\\
$\frac{(p_\Par-p_\Perp)}{p_0}\frac{B_xB_y}{B^2}$ & $0.198$ & $0.129$ & $0.0178 $ & $0.017$ & $0.654$\\
$\frac{4\pi(p_\Par-p_\Perp)}{B^2}$ & $-1.366$ & $0.51$ & $0.098$ & $-2.632$ & $-0.083$\\
$\frac{-B_xB_y}{(B^2/2)}$ & $0.5895$ & $0.1043$ & $0.0067$ & $0.3744$ & $0.8611$ \\
$\frac{\rho V_x \delta V_y}{(B^2/8\pi)}$ & $0.3323$ & $0.2725$ & $0.017$ & $-0.5307$ & $1.2704$ \\
$\frac{4\pi (p_\Par-p_\Perp)}{B^2} \frac{B_xB_y}{(B^2/2)}$ & $0.7356$
& $0.3718$ & $0.0714$ & $0.032$ & $1.807$ \\
$\frac{W_{xy}}{(B^2/8\pi)}$ & $1.6574$ & $0.6598$ & $0.084$ & $0.4364$ & $3.7159$ \\
$\frac{\alpha_R}{\alpha_M}$ & $0.5357$ & $0.3975$ & $0.024$ & $-0.9105$ & $2.084$ \\
$\frac{\alpha_A}{\alpha_M}$ & $1.2287$ & $0.5504$ & $0.119$ & $0.0854$ & $2.7243$ \\
$\frac{\rho}{\rho_0}$ & $0.99935$ & $2.3 \times 10^{-5}$ & $1.1 \times 10^{-5}$ & $0.9993$ & $0.9994$\\
$\frac{p_\Perp B_0}{B p_0}$ & $3.557$ & $1.665$ & $-^a$ & $1.1178$ & $7.929$\\
$\frac{p_\Par B^2 \rho_0^2}{\rho^2 B_0^2 p_0}$ 
   & $3.144 \times 10^3$ & $3.49 \times 10^3$
& $-$ & $585.4$ & $1.993 \times 10^4$\\
\enddata \tablenotetext{a} {We calculate the error using the
autocorrelation time only for quantities that saturate to a steady
state after 5 orbits. Estimate for correlation time $\tau_{int}$ is based on the 
discussion in \citet{nev05}.
$p_\Perp$ and $p_\Par$ show a secular growth
with time, so this way of expressing them as an average and an error
is not applicable.}
\end{deluxetable}

\begin{deluxetable}{ccccccc}
\tabletypesize{\scriptsize}
\tablecaption{Simulations with an Explicit Collision Term\label{tab:tab3}}
\tablewidth{0pt}
\tablehead{
\colhead{$\nu/\Omega$} & \colhead{$\langle \langle 4\pi \Delta p/B^2 \rangle \rangle$}
& \colhead{$ \langle \langle - \frac{B_x B_y}{4\pi p_0} \rangle \rangle $}
& \colhead{$(\langle \langle \frac{\rho V_x \delta V_y}{p_0} \rangle 
\rangle$}  & \colhead{$\langle \langle \frac{\Delta p^*}{B^2}\frac{B_xB_y}{p_0} 
\rangle \rangle$} & \colhead{$\alpha_A/\alpha_M$} & \colhead{$\alpha_A/\alpha_A(\nu=0)$} }
\startdata
$0$ & $-1.41$ & $0.18$ & $0.082$ & $0.196$ & $1.09$ & $1$ \\
$1$ & $-1.47$ & $0.152$ & $0.072$ & $0.173$ & $ 1.14$ & $0.88$ \\
$3$ & $-1.43$ & $0.178$ & $0.08 $ & $0.181$ & $1.02$ & $0.92$ \\
$10$ & $-1.35$ & $0.165$ & $0.071 $ & $0.159$ & $0.96$ & $0.81$ \\
$20$ & $-1.24$ & $0.174$ & $0.070 $ & $0.136$ & $0.78$ & $0.69$\\
$30$ & $-1.01$ & $0.213$ & $0.070 $ & $0.113$ & $0.53$ & $0.58$\\
$40$& $-0.87$ & $0.239$ & $0.070 $ & $0.095$ & $0.4$ & $0.48$\\
$100$ & $-0.43$ & $0.223$ & $0.06 $ & $0.032$ & $0.14$ & $0.16$\\
\enddata
\tablenotetext{*}{$\Delta p=(p_\Par-p_\Perp)$}
\end{deluxetable}

\end{document}